\documentstyle[12pt]{article}
\def\Tr#1{{{#1}}^{\rm tr}}  
\def\Long#1{{{#1}}^{\rm long}}

\def\Rom#1{\expandafter\uppercase\expandafter{\romannumeral #1}}
\let\fef=\ref
\def\ref#1{(\fef{#1})}
\makeatletter
\let\@rangle@=\rangle  
\let\@langle@=\langle  \let\@mid@=\mid 
\def\rangle{\ifinner\else\,\fi\@rangle@}
\def\langle{\@langle@\ifinner\else\,\fi}
\def\mid{\ifinner|\else\@mid@\fi}

\makeatother

\def\inv{\stackrel{R}{\longrightarrow}}

\def\vac#1{\langle0\mid#1\mid0\rangle}

\def\diag#1{\relax}
\begin{document}

\title{Conformally Invariant Green Functions of 
Current and Energy-Momentum
Tensor in Spaces of Even Dimension $D\ge 4$.}

\author{E.S.Fradkin\\
Institute for Advanced Study, Princeton,
NJ08540, USA\\
 {\it and \/} P.N. Lebedev Physical Institute,\\
 Leninsky 
Prospect, 53, 117924, Moscow, Russia.\\
\\
M.Ya.Palchik\\
Inst.\ of Automation and
Electrometry, Novosibirsk 630090, Russia}
\date{}

\maketitle
\vglue-5in
\hbox to\hsize{\hfill IASSNS-HEP-97/122}
\vglue5in


\bigskip


\begin{abstract}
We study the conformally invariant quantum field theory in spaces of
even dimension $D\ge 4$. The conformal transformations of current
$j_\mu$ and energy-momentum tensor $T_{\mu\nu}$ are examined.
It is shown that the set of conformal transformations of particular
kind corresponds to the canonical (unlike anomalous) dimensions $l_j=D-1$ 
and $l_T=D$ of those fields. 
These transformations cannot be derived by a smooth transiton
from anomalous dimensions.
The structure of representations
of the conformal group, which correspond to these canonical dimensions,
is analyzed, and new expressions for the propagators 
$\langle j_\mu j_\nu \rangle$ and $\langle T_{\mu\nu} T_{\rho\sigma}\rangle$ 
are derived. The latter expressions have integrable singularities.
It is shown that both propagators satisfy non-trivial Ward identities.
The higher Green functions of the fields $j_\mu$ and $T_{\mu\nu}$
are considered. The conformal QED and linear conformal gravity are
discussed. We obtain the expressions for invariant propagators of
electromagnetic and gravitational fields. The integrations over
internal photon and graviton lines are performed. The integrals
are shown to be conformally invariant and convergent, provided
that the new expressions for the propagators are used.
\end{abstract}

\newpage


\baselineskip=14pt 

\section{Introduction}

A family of exactly solvable conformal models in $D$ dimensions
is studied in~[1--5]. Notwithstanding that the conformal group
in $D\ge 3$ is finite-parametric, these models are in many
respects similar to two-dimensional conformal theories,
and coincide in $D=2$ with minimal models~[6--8]. The approach
presented in~[1--2] is based on conformally invariant~[3] Ward
identities for the conserved current $j_\mu$ and the energy-momentum
tensor $T_{\mu\nu}$. In $D$-dimensional case the latter fields 
have the canonical dimensions
\begin{equation}\label{1.1}
l_j=D-1,\qquad l_T=D.
\end{equation}
The conformal symmetry is supposed to be exact, and the energy-momentum
tensor to be traceless: $T_{\mu\nu}=0$. 

The conformally invariant propagators
\begin{equation}\label{1.2}
\langle j_\mu(x_1) j_\nu(x_2) \rangle, \qquad
\langle T_{\mu\nu}(x_1) T_{\rho\sigma}(x_2)\rangle
\end{equation}
are known to be ill-defined for even $D\ge 4$ due to formally
non-integrable factors $(x_{12}^2)^{-D+1}, (x_{12}^2)^{-D}$.
This leads to substantial difficulties when one attempts to include
gauge interactions in a conformally-invariant fashion. It is one
of the lasting problems in $D\ge 4$ conformal theory. The present
article is aimed to impart a detailed analysis and the solution
of this problem\footnote{We stress that the conformal symmetry
remains exact in this approach. Another approach was discussed lately
in~[9,10]. These works assume a special redefinition of conformal
propagators \protect\ref{1.2} which breaks the conformal symmetry.
The power factors are replaced by regularized expressions.
After that, the breakdown of the symmetry is shown to result in conformal
anomalies. Note that such effects might also be considered in our
approach, if the symmetry were to be broken.
 However, it falls beyond the focus of our discussion.}.

We shall show that for even $D\ge4$ the conformally invariant
propagators \ref{1.2} satisfy non-trivial Ward identities regardlessly
to the choice of regularization for their transversal parts. 
The r.h. sides of the Ward identities are non-zero due to contributions
of equal-time commutators between the components of current and energy-momentum
tensor. In conformal theory, each commutator is expressed through the
central charge $C_j$ or $C_T$.

As shown in~[1] (see also~[2,3]), the $D$-dimensional analogues of the
central charge may be $c$-number or, as well, operator valued. They
arise in operator product expansions
$$j_\mu(x) j_\nu(0) = [C_j]+[P_j(x)]+\ldots,$$
$$T_{\mu\nu}(x)T_{\rho\sigma}(0)= [C_T]+[P_T(x)]+\ldots,$$
where $P_j(x)$ and $P_T(x)$ are the scalar conformal fields of 
the dimension
$$d_P^j = d_P^T = D-2.$$
The operator analogues of the central charge, the fields $P_j(x)$ and
$P_T(x)$, contribute to the Ward identities for higher Green functions
$\langle j_\mu j_\nu\ldots\rangle$ and
$\langle T_{\mu\nu} T_{\rho\sigma}\ldots\rangle$, see~[3] for more details.

The propagators~\ref{1.2}, as well as higher Green functions,
may be chosen to have no longitudinal parts. Let us remind
that the expectation values of $T$-ordered products of the fields are
defined modulo quasilocal terms. One may employ this arbitrariness
to pass to transversal propagators. However, such a redefinition
of the propagators breaks the conformal symmetry. In the section~2
we show that longitudinal parts of Euclidean propagators
are uniquely fixed by conformal symmetry.

As an example, let us start with two-dimensional conformal models.
The problems due to divergencies are absent in this case.
Still, the Ward identities are non-trivial, as in the case of $D\ge4$.
The conformally invariant propagator of the current
\begin{equation}\label{1.3}
\left.\langle j_\mu(x) j_\nu(0)\rangle\right|_{D=2} = 
- \frac{1}{8\pi}C_j\left(\delta_{\mu\nu}-2\frac{x_\mu x_\nu}{x^2}
\right)\frac{1}{x^2} = - \frac{1}{4\pi} C_j
\partial_\mu\partial_\nu\ln x^2
\end{equation}
satisfies the following Ward identity:
\begin{equation}\label{1.4}
\partial_\mu\langle j_\mu(x)j_\nu(0)\rangle = C_j
\partial_\nu\delta(x).
\end{equation}
Consider the traceless energy-momentum tensor. The conformally
invariant propagator
\begin{equation}\label{1.5}
\left.\langle T_{\mu\nu}(x)T_{\rho\sigma}(0)\rangle\right|_{D=2} = 
\frac{1}{8\pi}C_T\left.\left[
g_{\mu\rho}(x)g_{\nu\sigma}(x)+g_{\mu\sigma}(x)g_{\nu\rho}(x)-
\delta_{\mu\nu}\delta_{\rho\sigma}\right]\frac{1}{(x^2)^{2+\epsilon}}
\right|_{\epsilon=0},
\end{equation}
where $g_{\mu\nu}(x)=\delta_{\mu\nu}-2\frac{x_\mu x_\nu}{x^2}$,
is finite when $\epsilon\to 0$. Using the identity for $D=2$
\begin{eqnarray} \label{1.6}
&& \delta_{\mu\rho}\partial_\nu\partial_\sigma+
\delta_{\mu\sigma}\partial_\nu\partial_\rho+
\delta_{\nu\rho}\partial_\mu\partial_\sigma+
\delta_{\nu\sigma}\partial_\mu\partial_\rho\nonumber \\
&&\quad{}=2\left(\delta_{\mu\nu}\partial_\rho\partial_\sigma+
\delta_{\rho\sigma}\partial_\mu\partial_\nu\right)+
\left(\delta_{\mu\rho}\delta_{\nu\sigma}+\delta_{\mu\sigma}\delta_{\nu\rho}
-\delta_{\mu\nu}\delta_{\rho\sigma}\right)\Box,
\end{eqnarray}
one can bring it to the form
\begin{eqnarray} \label{1.7}
&&\left. \langle T_{\mu\nu}(x)T_{\rho\sigma}(0)\rangle\right|_{D=2}=
-\frac{C_T}{24}\bigl[\left(\partial_\mu\partial_\nu-\frac{1}{2}
\delta_{\mu\nu}\Box\right)\left(\partial_\rho\partial_\sigma-\frac{1}{2}
\delta_{\rho\sigma}\Box\right)\nonumber \\
&& \quad{}-\frac{1}{8}\left(\delta_{\mu\rho}\delta_{\nu\sigma}+
\delta_{\mu\sigma}\delta_{\nu\rho}-\delta_{\mu\nu}\delta_{\rho\sigma}
\right)\Box^2\bigr]\frac{1}{\Box}\delta(x),
\end{eqnarray}
where $\frac{1}{\Box}\delta(x)=-\frac{1}{4\pi}\ln x^2$. The
Ward identity may be derived directly from~\ref{1.5}, conducting
the calculations for $\epsilon\ne 0$ and then passing to the limit~[11]
$\lim_{\epsilon=0}\left[\epsilon\frac{1}{(x^2)^{1+\epsilon}}\right]=
\pi\delta(x)$. As the result one obtains
\begin{eqnarray} \label{1.8}
&& \partial_\nu\left.\langle T_{\mu\nu}(x)T_{\rho\sigma}(0)\right|_{D=2}=
-\frac{C_T}{24}\bigl[\partial_\mu\partial_\rho\partial_\sigma-
\frac{1}{4}\left(\delta_{\mu\rho}\partial_\sigma+\delta_{\mu\sigma}
\partial_\rho\right)\Box\nonumber \\
&& \quad\quad\quad\quad{}-\frac{1}{4}\delta_{\rho\sigma}\partial_\mu
\Box\bigr]\delta(x).
\end{eqnarray}
This equation is a particular case of the Ward identity to be found
in the next section for the case of any even $D$. Passing in  \ref{1.5},
\ref{1.8} to complex variables
\begin{eqnarray*}
&& x^\pm = x_1 \pm i x_2,\quad
\partial_\pm = \frac{1}{2}\left(\partial_1 \mp i\partial_2\right),\quad
x^+\to z,\quad x^- \to \bar{z},
\partial_+\to\partial_z,\quad
\partial_-\to\partial_{\bar{z}} \\
&& T_{zz}\sim\frac{1}{2}(T_{11}-T_{22}-iT_{12}-iT_{21}),\quad
T_{z\bar{z}}\sim\frac{1}{2}(T_{11}+T_{22}),
\end{eqnarray*}
we find well-known results
\begin{eqnarray*}
&&\langle T_{zz}(z)T_{zz}(0)=\frac{1}{6}\pi C_T
(\partial_z)^4(\partial_z\partial_{\bar{z}})^{-1}
\delta^{(2)}(z,\bar{z})\sim\frac{1}{z^4},\\
&&\partial_{\bar{z}}\langle T_{zz}(z)T_{zz}(0)=\frac{\pi C_T}{6}
(\partial_z)^3 \delta^{(2)}(z,\bar{z}).
\end{eqnarray*}
The conformally invariant expression \ref{1.7} differs from the
transversal propagator 
\begin{equation}\label{1.9}
\sim (\partial_\mu\partial_\nu-\delta_{\mu\nu}\Box)
(\partial_\rho\partial_\sigma-\delta_{\rho\sigma}\Box)\ln x^2
\end{equation}
by quasilocal terms.
These terms are present in \ref{1.7} and may be eliminated by a redefinition 
of the $T$-ordered product of $T_{\mu\nu}$ components. However the latter
will lead to a breakdown of conformal invariance, and to non-zero trace of
the energy-momentum tensor.

The expression for the propagator $\langle TT\rangle$ for $D>2$
results from \ref{1.5} after the change $(x^2)^{-2-\epsilon} \to
(x^2)^{-D-\epsilon}$. It is divergent in the limit $\epsilon\to0$ for
even $D\ge 4$, since $\left.(x^2)^{-D-\epsilon}\right|_{\epsilon\to0}\sim 
(1/\epsilon)\Box^{D/2}\delta(x)$. Similarly, the expression for the
propagator $\langle jj\rangle$ for $D>2$ 
results from \ref{1.3} after the change $(x^2)^{-1} \to
(x^2)^{-D+1+\epsilon}$, and also diverges in the limit $\epsilon\to0$.
Note that in two-dimensional case these divergences are cancelled; it
can be seen from the differential representations \ref{1.9},\ref{1.5}.
We shall show in sections~3--5 that for $D\ge 4$ the latter divergences
are formal and contribute neither to Feynman graphs nor to any other 
contractions available in conformal theory. The presence of the above
divergences is caused by peculiar properties of conformal transformations
for the fields of dimensions~\ref{1.1}. We shall demonstrate that
the conformal transformations of the fields, as well as invariant averages
of the fields with anomalous dimensions
\begin{equation}\label{1.10}
l=l_j+\epsilon,\quad l=l_T+\epsilon,
\end{equation}
can not be analytically continued to the values at $\epsilon=0$ because
the representations of the conformal group have singular properties at these 
points. These representations belong to a series of exceptional integer
points~[12,13] and must be reconstructed completely. This is done in
sections~3,4. Conformal transformations of the fields $j_\mu$, $T_{\mu\nu}$ 
for $\epsilon=0$ will be shown to have different structures in longitudinal 
and transversal sectors. Correspondingly, the two types of invariant
kernels contributing to the propagators~\ref{1.2} will be shown to exist.
The latter leads to a specific structure of the conformal fields
$j_\mu(x)$ and $T_{\mu\nu}(x)$. Two mutually orthogonal sectors of the
Hilbert space, having different physical meanings~[1,2] (see also Sec.6), 
correspond to each of these fields. The elements of the first sector are the 
equivalence classes, each including a whole set of states. The 
transformations inside an equivalence class do not alter physical results.
We shall show that the above divergences may be eliminated using the 
transformations inside equivalence classes and hence do not effect the 
results, see Sec.~3,4.

In sections~5,6 the conformal QED and the linear conformal gravity are 
considered. The problem of how to single out the contribution due
to gauge interactions into the conformally invariant Gauge functions
\begin{equation}\label{1.11}
\langle j_\mu\ldots\rangle,\quad\langle T_{\mu\nu}\ldots\rangle,
\end{equation}
where dots stand for any set of fields, will also be discussed.
The above Green functions can be found~[1,2] from Ward identities up to 
transversal conformal parts solely caused by gauge interactions.
The framework developed here allows to single out this contribution
{\bf uniquely,\/} see also~[4,5]. The Green functions~\ref{1.11} are 
calculated from Ward identities, their remaining parts describe a theory 
of direct (non-gauge) interaction. As a result, in a theory without gauge 
interactions the Green functions \ref{1.11} are uniquely determined by Ward 
identities. In our opinion, this result appears to be sufficiently 
promising, since it allows one to derive~[1,2] the $D$-dimensional analogues 
of minimal models. The additional conditions imposed on the 
functions~\ref{1.11}, which ensure the absence of contributions due to gauge 
interactions, as well as the number of models in $D>2$ and the transition to 
$D=2$ theory in this approach, are dealt with in~[4,5]. Here we present a 
more detailed justification of these conditions (section~6).

Let us remark that for the sake of completeness the section~3 contains a 
brief description of representations of the conformal group in integer 
points, to the length required for understanding the sections~4--6. At the 
first reading, section~3 may be skipped. A comprehensive analysis
of representations regarding 
conformal gauge theories is given in~[1] (see also~[14--17]).

\section{Ward Identities for the Propagators 
\setcounter{equation}{0}
$\langle j_\mu j_\nu \rangle$ and $\langle T_{\mu\nu} T_{\rho\sigma}\rangle$}

\subsection{Longitudinal Part of the Propagator of the Current}
Let $j_\mu^l(x)$ be a conformal quantum field of anomalous dimension $l$ in 
$D$-dimensional Euclidean space. The transformation of conformal inversion
\begin{equation}\label{2.1}
x_\mu \to R x_\mu = \frac{x_\mu}{x^2}
\end{equation}
induces the following transformation of the field $j_\mu^l$:
\begin{equation}\label{2.2}
j_\mu^l(x) \inv \frac{1}{(x^2)^l}g_{\mu\nu}(x)j_\nu(Rx),
\end{equation}
where
\begin{equation}\label{2.3}
g_{\mu\nu}(x) = \delta_{\mu\nu}-2 \frac{x_\mu x_\nu}{x^2}.
\end{equation}
The invariant propagator is determined from the condition
\begin{equation}\label{2.4}
\langle j_\mu^l(x_1) j_\nu^l(x_2)\rangle =
\frac{1}{(x_1^2)^l}
\frac{1}{(x_2^2)^l}
g_{\mu\rho}(x_1)g_{\nu\sigma}(x_2)\langle j_\rho(Rx_1)
j_\sigma(Rx_2)\rangle.
\end{equation}
The solution has the form
\begin{equation}\label{2.5}
\langle j_\mu^l(x_1) j_\nu^l(x_2)\rangle =
\frac{\tilde{C}_j}{(x_{12}^2)^l}g_{\mu\nu}(x_{12}),
\end{equation}
where $\tilde{C}_j$ is some constant.

Consider the behaviour of this propagator near the canonical dimension point
\begin{equation}\label{2.6}
l=l_j+\epsilon=D-1+\epsilon.
\end{equation}
All its contractions of the type
\begin{equation}\label{2.7}
\left.\int dx_1\,dx_2\, A_\mu^{D-l}(x_1)
\langle j_\mu^l(x_1) j_\nu^l(x_2)\rangle A_\nu^{D-l}(x_2)
\right|_{\epsilon\to0},
\end{equation}
where $A_\mu^{D-l}$ is a conformal field of dimension $D-l$, diverge in the 
limit of $\epsilon=0$. The term leading in $\epsilon$ is calculated using the 
relations~[11]
\begin{equation}\label{2.8}
\left.\frac{1}{(x^2)^{D/2+k+\epsilon}}\right|_{\epsilon\to0}\simeq
\frac{1}{\epsilon}\frac{\pi^{D/2}4^{-k}}{
\Gamma\left(\frac{D}{2}+k\right)\Gamma\left(k+1\right)}\Box^k\delta(x).
\end{equation}
Let us rewrite the propagator as
\begin{eqnarray} \label{2.9}
&& \Delta_{\mu\nu}^\epsilon(x_{12})=
\langle j_\mu^l(x_1) j_\nu^l(x_2)\rangle=
{\tilde{C}_j\over2(D-1+\epsilon)(D-2+\epsilon)}\nonumber\\
&&{}\times\left[
(\delta_{\mu\nu}\Box-\partial_\mu\partial_\nu)
\frac{1}{(x_{12}^2)^{D-2+\epsilon}} -
\frac{\epsilon}{D-2+2\epsilon}\delta_{\mu\nu}\Box
\frac{1}{(x_{12}^2)^{D-2+\epsilon}}\right].
\end{eqnarray}
Hence it is clear that the divergent part of the propagator is transversal.

Note that in the limit $\epsilon=0$ the field $A_\mu^{D-l}$ coincides with 
the electromagnetic field. A conformally invariant regularization 
$(l_A=1\to l_A^\epsilon=1-\epsilon)$ of the field $A_\mu(x)$ will be treated 
in section~5. The contractions~\ref{2.7} in the longitudinal sector
\begin{equation}\label{2.10}
\left.\Long{A}_\mu(x)\right|_{\epsilon=0}=\partial_\mu\varphi,\quad
\left.\Tr{A}_\mu(x)\right|_{\epsilon=0}=0
\end{equation}
are finite in virtue of~\ref{2.9}. As will be shown in sections~4,5,
a new expression for the propagator $\langle j_\mu j_\nu\rangle$ appears in 
the transversal sector, while the kernel~\ref{2.9} may participate in the
longitudinal sector only. Below in this section we consider a theory without 
electromagnetic interaction. By definition, only the contractions~\ref{2.7} 
with longitudinal fields~\ref{2.10}
\begin{equation}\label{2.11}
\left.\int dx_1\, dx_2\, \Long{A}_\mu(x_1)
\Delta_{\mu\nu}^\epsilon(x_{12})\Long{A}_\nu(x_2)\right|_{\epsilon=0}
\end{equation}
are present in that theory. These contractions are transversal because the 
divergent part of the kernel~\ref{2.9} does not contribute to them.

Consider a Ward identity for the propagator $\langle j_\mu j_\nu\rangle$.
Define the l.h.s.\ of the identity by the relation
$$\partial_\mu
\langle j_\mu(x_1) j_\nu(x_2)\rangle=\lim_{\epsilon\to0}
\partial_\mu
\langle j_\mu^l(x_1) j_\nu^l(x_2)\rangle.$$
From~\ref{2.5} one has
$$\partial_\mu\langle j_\mu^l(x_1) j_\nu^l(x_2)\rangle=
-\frac{\epsilon\tilde{C}_j}{(D-1+\epsilon)}\partial_\nu^{x_1}
\frac{1}{(x_{12})^{D-1+\epsilon}}.$$
Using~\ref{2.8} for $k=\frac{D-2}{2}$
$$\lim_{\epsilon=0}\epsilon\frac{1}{(x_{12})^{D-1+\epsilon}} =
\frac{4^{-\frac{D-2}{2}}\pi^{D/2}}{\Gamma\left(D-1\right)\Gamma\left(
\frac{D}{2}\right)}\Box^{\frac{D-2}{2}}\delta(x),$$
we find
\begin{equation}\label{2.12}
\partial_\mu\langle j_\mu(x_1) j_\nu(x_2)\rangle=
C_j\partial_\nu\Box^{\frac{D-2}{2}}\delta(x_{12}),
\end{equation}
where
$$C_j = -\frac{4^{-\frac{D-2}{2}}\pi^{D/2}}{\Gamma\left(\frac{D}{2}\right)
\Gamma\left(D\right)}\tilde{C}_j.$$
One easily checks that this identity is conformally invariant. The l.h.s\ 
of~\ref{2.12} transforms as a Euclidean average
\begin{equation}\label{2.13}
\langle j(x_1)j_\nu(x_2)\rangle\ne 0,
\end{equation}
where $j(x)=\partial_\mu j_\mu(x)$ is a conformal scalar of dimension $D$:
$$j(x) \to \frac{1}{(x^2)^D}j(Rx).$$
This transformation law is proved with the help of identities
\begin{equation}\label{2.14}
\partial_\mu^x = \frac{1}{x^2}g_{\mu\nu}(x)\partial_\nu^{Rx},\
g_{\mu\rho}(x)g_{\rho\nu}(x)=\delta_{\mu\nu}, \
\partial_\mu\left[\frac{1}{(x^2)^{D-1}}g_{\mu\nu}(x)\right]=0.
\end{equation}
Note that the conformally invariant average
including scalar and vector vanishes in the general case
$$\langle j_\mu^l(x_1) j_\nu^l(x_2)\rangle
=0,\quad\hbox{if $l_1\ne D$ and $l_2\ne D-1$}$$
for any dimension save
$$l_1=D,\quad l_2=D-1,$$
the latter are exceptional ones. The r.h.s.\ of the identity~\ref{2.12} also 
transforms like the quantity~\ref{2.13} --- this may be checked with the help 
of relation
\begin{equation}\label{2.15}
\delta(Rx_{12})=\left(x_{1}^2\right)^D\delta(x_{12}).
\end{equation}
The invariance of the identity~\ref{2.12} is proved.

Thus, the expansion in $\epsilon$ of the regularized propagator
$\Delta_{\mu\nu}^\epsilon$ has a finite longitudinal term
\begin{equation}\label{2.16}
\Long{\Delta}_{\mu\nu}(x_{12})=\langle j_\mu(x_1)j_\nu(x_2)\rangle^{\rm long}
= C_j\partial_\mu\partial_\nu\Box^{\frac{D-4}{2}}\delta(x_{12}).
\end{equation}
In the discussions below, the following form of the regularized
propagator is useful
\begin{eqnarray} \label{2.17}
&& \Delta_{\mu\nu}^\epsilon=
\frac{\tilde{C}_j}{2(D-1+\epsilon)(D-2+2\epsilon)}
\left(\delta_{\mu\nu}\Box-\partial_\mu\partial_\nu\right)
\frac{1}{(x_{12}^2)^{D-2+\epsilon}}\nonumber\\
&&\qquad\qquad\quad{}+C_j\partial_\mu\partial_\nu
\Box^{\frac{D-4}{2}}\delta(x_{12})+O(\epsilon).
\end{eqnarray}
Note that in the derivation of the regularized expression~\ref{2.17} in some 
publications the finite longitudinal term was missing, and the propagator 
was identified with the regularized transversal part of this expression.

\subsection{Transversal Part of the Propagator of the 
Energy-Mo\-me\-n\-tum Tensor}
Let $T_{\mu\nu}^l(x)$ be a traceless symmetric tensor of dimension $l$.
Under conformal inversion it transforms as
\begin{equation}\label{2.18}
T_{\mu\nu}^l(x) \inv \frac{1}{(x^2)^l} g_{\mu\rho}(x)
g_{\nu\sigma}(x) T_{\rho\sigma}^l(Rx).
\end{equation}
The conformal propagator is determined by the condition of invariance
\begin{eqnarray} \label{2.19}
&& \langle T_{\mu\nu}^l(x_1) T_{\rho\sigma}^l(x_2) \rangle
=\frac{1}{(x_1^2x_2^2)^l}g_{\mu\alpha}(x_1)g_{\nu\beta}(x_1)
g_{\rho\lambda}(x_2)g_{\sigma\tau}(x_2) \nonumber \\
&&\qquad\qquad\quad{}\times
\langle T_{\alpha\beta}^l(Rx_1)T_{\lambda\tau}^l(Rx_2)\rangle
\end{eqnarray}
and has the form
\begin{equation}\label{2.20}
\langle T_{\mu\nu}^l(x_1) T_{\rho\sigma}^l(x_2) \rangle
=\tilde{C}_T\left[g_{\mu\rho}(x_{12})g_{\nu\sigma}(x_{12})+
g_{\mu\sigma}(x_{12})g_{\nu\rho}(x_{12})-\frac{2}{D}
\delta_{\mu\nu}\delta_{\rho\sigma}\right]
\frac{1}{(x_{12}^2)^l},
\end{equation}
where $\tilde{C}_T$ is some constant. Consider its behaviour near the
point of canonical dimension $l_T$:
\begin{equation}\label{2.21}
l=l_T+\epsilon=D+\epsilon.
\end{equation}
The contractions
\begin{equation}\label{2.22}
\left.\int dx_1\, dx_2\, h_{\mu\nu}^{D-l}(x_1)
\langle T_{\mu\nu}^l(x_1) T_{\rho\sigma}^l(x_2) \rangle
h_{\rho\sigma}^{D-l}(x_2)\right|_{\epsilon\to0}
\end{equation}
are divergent in the limit of $\epsilon=0$ if $\Tr{h}_{\mu\nu}(x)\ne0$.
One can check that the singular in $\epsilon$ part of this propagator is
transversal and has the form:
\begin{equation}\label{2.23}
\Delta_{\mu\nu}(x_{12}) \sim \tilde{C}_T
\Tr{H}_{\mu\nu}\left(\frac{\partial}{\partial x}\right)
\frac{1}{(x_{12}^2)^{D-2+\epsilon}}+O(1),
\end{equation}
where $\Tr{H}_{\mu\nu}$ is the transversal (in each index) differential
operator
\begin{eqnarray} \label{224}
&& \partial_\mu\Tr{H}_{\mu\nu\rho\sigma}
\left(\frac{\partial}{\partial x}\right) = 0,\quad
\Tr{H}_{\mu\nu\rho\sigma}=\Tr{H}_{\rho\sigma\mu\nu}=
\Tr{H}_{\nu\mu\rho\sigma},\nonumber\\
&& \Tr{H}_{\mu\nu\rho\sigma}
\left(\frac{\partial}{\partial x}\right) =
\biggl\{\frac{D-2}{D-1}\partial_\mu\partial_\nu\partial_\rho\partial_\sigma
-\frac{1}{2}\bigl(\delta_{\mu\rho}\partial_\nu\partial_\sigma+
\delta_{\mu\sigma}\partial_\nu\partial_\rho \nonumber \\
&&{}+\delta_{\nu\rho}\partial_\mu\partial_\sigma+
\delta_{\nu\sigma}\partial_\mu\partial_\rho\bigr)\Box-
\frac{1}{(D-1)}\bigl(\delta_{\mu\nu}
\partial_\rho\partial_\sigma+\delta_{\rho\sigma}\partial_\mu\partial_\nu
\bigr)\Box\nonumber \\
&&\quad\quad{}+\frac{1}{2}\bigl(\delta_{\mu\rho}\delta_{\nu\sigma}+
\delta_{\mu\sigma}\delta_{\nu\rho}\bigr)\Box^2-
\frac{1}{(D-1)}\delta_{\mu\nu}\delta_{\rho\sigma}
\Box^2\biggr\}.
\end{eqnarray}
Thus the divergent part of the contraction~\ref{2.22} is caused by a
contribution of the transversal component of the field $h_{\mu\nu}(x)$ when
$\epsilon\to0$.

In the limit $\epsilon=0$ the field $h_{\mu\nu}(x)$ coincides with the
traceless part of metric tensor, which has a zero dimension: $l_h=
D-l_T=0$. In section~5 we describe a conformally invariant regularization of 
the field $h_{\mu\nu}$ ($l_h\to l_h^\epsilon=-\epsilon$). In the longitudinal
sector consisting of the fields
\begin{equation}\label{225}
\left.\Long{h}_{\mu\nu}(x)\right|_{\epsilon=0}=
\partial_\mu h_\nu(x)+\partial_\nu h_\mu(x) -\frac{2}{D}
\delta_{\mu\nu}\partial_\lambda h_\lambda(x),\quad
\left.\Tr{h}_{\mu\nu}(x)\right|_{\epsilon=0}=0
\end{equation}
we get the contractions which are finite due to~\ref{2.23}
\begin{equation}\label{2.24}
\int dx_1\, dx_2\,
\Long{h}_{\mu\nu}(x_1)
\langle T_{\mu\nu}^l(x_1) T_{\rho\sigma}^l(x_2) \rangle
\Long{h}_{\rho\sigma}(x_2).
\end{equation}
In the theory with no gravitational interaction only the 
contractions~\ref{2.24} may appear. We show in sections~4--5 that in the 
transversal sector of the conformal gravity a new expression for the 
propagator $\langle T_{\mu\nu}T_{\rho\sigma}\rangle$ arises. This expression 
is free of divergences at $\epsilon=0$.

Consider the Ward identity for the propagator. By definition, we set
$$\partial_\nu
\langle T_{\mu\nu}(x_1) T_{\rho\sigma}(x_2) \rangle=\lim_{\epsilon=0}
\partial_\nu\langle T_{\mu\nu}^l(x_1) T_{\rho\sigma}^l(x_2) \rangle.$$
The divergent part present in~\ref{2.23} vanishes after taking the 
derivative. As the result we have from~\ref{2.20}
\begin{eqnarray} \label{2.25}
&& \partial_\nu\langle T_{\mu\nu}^l(x_1) T_{\rho\sigma}^l(x_2) \rangle=
\frac{\epsilon\tilde{C}_T}{(D-1+\epsilon)(D+1+\epsilon)}
\biggl[\partial_\mu\partial_\rho\partial_\sigma\nonumber \\
&&{}-\frac{(D-1+\epsilon)}{2(D+2\epsilon)}
\left(\delta_{\mu\rho}\partial_\sigma+\delta_{\mu\sigma}\partial_\rho\right)
\Box-\frac{1+\epsilon}{D(D+\epsilon)}\delta_{\rho\sigma}
\partial_\mu\Box\biggr]\frac{1}{(x_{12}^2)^{D-1+\epsilon}}.
\end{eqnarray}
Using the relation~\ref{2.8} for $k=\frac{D-2}{2}$ one gets
\begin{eqnarray} \label{2.26}
&& \partial_\nu\langle T_{\mu\nu}(x_1) T_{\rho\sigma}(x_2) \rangle=
C_T\biggl[\partial_\mu\partial_\rho\partial_\sigma-
\frac{D-1}{2D}
\left(\delta_{\mu\rho}\partial_\sigma+\delta_{\mu\sigma}\partial_\rho\right)
\Box\nonumber\\
&&\qquad\qquad\quad{}-\frac{1}{D^2} 
\delta_{\rho\sigma}\partial_\mu\Box\biggr]
\Box^{\frac{D-2}{2}}\delta(x_{12}),
\end{eqnarray}
where 
$$C_T=-\tilde{C}_T\pi^{D/2}4^{-\frac{D-2}{2}}
\left[\Gamma\left(D+2\right)\Gamma\left(\frac{D}{2}\right)\right]^{-1}.$$
For $D=2$ this identity becomes naturally~\ref{1.8}.

It is not hard to show that the identity~\ref{2.26} is conformally invariant. 
The  l.h.s.\ transforms as a conformally invariant average of a vector and
a tensor:
\begin{equation}\label{2.27}
\langle T_\mu(x_1)T_{\rho\sigma}(x_2)\rangle\inv
\frac{1}{(x^2_1)^{D+1}}\frac{1}{(x^2_2)^D}
g_{\mu\nu}(x_1)g_{\rho\alpha}(x_2)g_{\sigma\beta}(x_2)
\langle T_\nu(Rx_1) T_{\alpha\beta}(Rx_2)\rangle,
\end{equation}
where $T_\mu(x)=\partial_\nu T_{\mu\nu}(x)$ is a conformal vector of 
dimension $d+1$. One easily checks that for $l=D$
the quantity $\partial_\nu T_{\mu\nu}(x)$ is indeed the conformal vector
$$\partial_\nu^x T_{\mu\nu}(x) \inv
\frac{1}{(x^2)^{D+1}}
g_{\mu\rho}(x)\partial_\nu^{Rx} T_{\rho\nu}(Rx).$$
Notice that in the general case the conformally invariant average
of a vector and a tensor is zero
$$\langle T_\mu^{l_1}(x_1)
T_{\rho\sigma}^{l_2}(x_2)\rangle,\quad
\hbox{if $l_1\ne D+1$ and $l_2\ne D$,}$$
for any dimension except
$$l_1=D+1,\quad l_2=D.$$
These dimensions are exceptional, and $\langle T_\mu T_{\rho\sigma}
\rangle\ne0$. The r.h.s.\ of the identity~\ref{2.26} also transforms by the
law~\ref{2.27}. This is checked with the help of~\ref{2.15}. The invariance 
of the identity is proved.

>From the discussion above follows that the expansion~\ref{2.23} includes
a finite longitudinal term of the type
\begin{equation}\label{2.28}
\Long{\Delta}_{\mu\nu\rho\sigma}(x_{12})=
\langle T_{\mu\nu}(x_1)T_{\rho\sigma}(x_2)\rangle^{\rm long}=
\partial_\mu\Delta_{\nu\rho\sigma}(x_{12})+
\partial_\nu\Delta_{\mu\rho\sigma}(x_{12}) -\frac{2}{D}
\delta_{\mu\nu}\partial_\lambda\Delta_{\lambda\rho\sigma}(x_{12}).
\end{equation}
One finds from the Ward identity~\ref{2.26}:
\begin{eqnarray} \label{2.29}
&& \Delta_{\mu\rho\sigma}(x_{12}) = C_T \biggl\{
\frac{(2D^2-3D+2)}{2D(D-1)}\partial_\mu\partial_\rho\partial_\sigma
\nonumber \\
&& \quad{}-\frac{D-1}{2D}
\left(\delta_{\mu\rho}\partial_\sigma+\delta_{\mu\sigma}\partial_\rho\right)
\Box-\frac{1}{2D(D-1)}\delta_{\rho\sigma}\partial_\mu\Box\biggr\}
\Box^{\frac{D-4}{2}}\delta(x_{12}).
\end{eqnarray}
Rewrite the expansion~\ref{2.23} in the form
\begin{eqnarray} \label{2.30}
&& \Delta_{\mu\nu\rho\sigma}^\epsilon(x_{12}) =\left.
\langle T_{\mu\nu}^l(x_1)T_{\rho\sigma}^l(x_2)\rangle\right|_{\epsilon\to0}
\nonumber \\
&& \simeq \Delta_{\mu\nu\rho\sigma}^{\epsilon\,\rm tr}(x_{12})+
 \Delta_{\mu\nu\rho\sigma}^{\rm long}(x_{12})+O(\epsilon),
\end{eqnarray}
where
$$ \Delta_{\mu\nu\rho\sigma}^{\epsilon\,\rm tr}(x_{12})\sim
\tilde{C}_T\Tr{H}_{\mu\nu\rho\sigma}
\left(\frac{\partial}{\partial x}\right)
\frac{1}{(x_{12}^2)^{D-2+\epsilon}}.$$
In a number of works dealing with the expansion~\ref{2.30}, the finite 
longitudinal term~\ref{2.28} is missing.

For what follows it is helpful to introduce projection operators~[1] 
selecting out longitudinal and transversal sectors. Let us start with the 
longitudinal projector. We put
\begin{eqnarray} \label{2.32}
&& P_{\mu\nu\rho}
\left(\frac{\partial}{\partial x}\right)
=-\frac{1}{2}\biggl[
\frac{(D-2)}{(D-1)}
\partial_\mu\partial_\nu\partial_\rho\frac{1}{\Box^2}\nonumber \\
&&\qquad\quad{} -
\left(\delta_{\mu\nu}\partial_\rho+\delta_{\mu\rho}\partial_\nu\right)
\frac{1}{\Box}+\frac{1}{D-1}\delta_{\rho\sigma}\partial_\mu\frac{1}{\Box}
\biggr].
\end{eqnarray}
It is not hard to check that the operator
\begin{equation}\label{2.33}
\Long{P}_{\mu\nu\rho\sigma}
\left(\frac{\partial}{\partial x}\right)
=\partial_\mu P_{\nu\rho\sigma}
\left(\frac{\partial}{\partial x}\right)
+\partial_\nu P_{\mu\rho\sigma}
\left(\frac{\partial}{\partial x}\right)
-\frac{2}{D}\delta_{\mu\nu}\partial_\lambda P_{\lambda\rho\sigma}
\left(\frac{\partial}{\partial x}\right)
\end{equation}
is a projection operator:
$$\Long{P}_{\mu\nu\rho\sigma}\Long{P}_{\rho\sigma\lambda\tau}=
\Long{P}_{\mu\nu\lambda\tau}.$$
The transversal projector is defined as
\begin{equation}\label{2.34}
\Tr{P}_{\mu\nu\rho\sigma}
\left(\frac{\partial}{\partial x}\right)
=I_{\mu\nu\rho\sigma}-\Long{P}_{\mu\nu\rho\sigma}
\left(\frac{\partial}{\partial x}\right),
\end{equation}
where
$$I_{\mu\nu\rho\sigma} =\frac{1}{2}\left(
\delta_{\mu\rho}\delta_{\nu\sigma}+
\delta_{\mu\sigma}\delta_{\nu\rho}-
\frac{2}{D}\delta_{\mu\nu}\delta_{\rho\sigma}\right).$$
One can show that this operator is related to the operator 
$\Tr{H}_{\mu\nu\rho\sigma}$ introduced above by the equality
\begin{equation}\label{2.35}
\Tr{P}_{\mu\nu\rho\sigma}
\left(\frac{\partial}{\partial x}\right) = 
\frac{1}{\Box^2}
\Tr{H}_{\mu\nu\rho\sigma}
\left(\frac{\partial}{\partial x}\right),
\end{equation}
where $\Tr{H}$ is given by the expression~\ref{224}.

\section{The Transformations of Conformal Group in Integer Points}
\setcounter{equation}{0}
In this section we demonstrate that the divergences of the 
expressions~\ref{2.17} and~\ref{2.30} at $\epsilon=0$ are caused by the
specific structure of representations of the conformal group, which are given 
by the transformation laws~\ref{2.2} and~\ref{2.18} for $l=l_j$ and $l=l_T$. 
Let us remind that the propagators which are invariant under the 
transformations of a certain symmetry group may be found as the kernels of 
invariant bilinear forms in the representation space. For example, the 
propagator of the irreducible conformal field $j_\mu^l$ may be introduced as 
the kernel of the form
\begin{equation}\label{3.1}
(A^{D-l},A^{D-l})=
\int dx_1\, dx_2\,
A_\mu^{D-l}(x_1)\Delta_{\mu\nu}^l(x_{12})A_\nu^{D-l}(x_2),
\end{equation} 
where $A_\mu^{D-l}$ are conformal fields of dimension $\tilde{l}=D-l$.
(We also remind that representations with dimensions $l$ and $\tilde{l}=
D-l$ are equivalent). Taking into account that $d^Dx=(x^2)^D
d^D(Rx)$, one obtains the expression~\ref{2.4} for the kernel 
$\Delta_{\mu\nu}^l(x)$, so that the kernel may be identified with the 
invariant propagator of the field $j_\mu^l$:
\begin{equation}\label{3.2}
\langle j_\mu^l(x_1) j_\nu^l(x_2) \rangle 
\sim \Delta_{\mu\nu}^l(x_{12}).
\end{equation} 
The latter is valid for either irreducible representations or their direct 
sums, since corresponding kernels are non-degenerate.

It is known that~[10,11] the representations of the conformal group defined 
by the transformation laws~\ref{2.2} and~\ref{2.18} are irreducible for all 
values of $l$ except the points
\begin{equation}\label{3.3}
l_j=D-1,\ l_A=D-l_j=1\quad\hbox{and}\quad
l_T=D,\ l_h=D-l_T=0,
\end{equation}
which belong to a series of exceptional integer points. The representations 
in these points are undecomposable, while the invariant kernels are 
degenerate. To derive invariant propagators of the fields with such 
dimensions, it is primarily necessary to construct irreducible 
representations and their direct sums. After that, the propagators may be 
determined from the relations of the type~\ref{3.2}. Let us demonstrate this 
on an example of the current.

\subsection{Irreducible Representations for the Fields $A_\mu$ and $j_\mu$}
The propagator of the current
\begin{equation}\label{3.4}
\Delta_{\mu\nu}(x_{12}) = 
\langle j_\mu(x_1)j_\nu(x_2)\rangle
\end{equation}
is the kernel of the invariant contraction
\begin{equation}\label{3.5}
(A,A) = 
\int dx_1\, dx_2\,
A_\mu(x_1)
\langle j_\mu(x_1)j_\nu(x_2)\rangle
A_\nu(x_2),
\end{equation}
where $A_\mu(x)$ is the electromagnetic potential. Consider the 
transformation law
\begin{equation}\label{3.6}
A_\mu(x) \inv A_\mu'(x) = U_R^A A_\mu(x) =\frac{1}{x^2}
g_{\mu\nu}(x)A_\mu(Rx).
\end{equation}
This transformation law leads to a number of difficulties. The 
contraction~\ref{3.5} is invariant if the propagator~\ref{3.4} satisfies the 
equation~\ref{2.4} after the formal transition $\epsilon=l-D+1\to0$ is
performed in the latter equation. Also, we know that the solution~\ref{2.5} 
of this equation for $\epsilon\ne 0$ does not admit such a transition to this 
limit for even $D\ge4$. Note however, that the equation~\ref{2.4} for $l=D-1$ 
admits (for even $D\ge4$) a special solution which cannot be obtained as a 
result of transition to the limit $\epsilon\to0$. This solution is 
transversal:
\begin{equation}\label{3.7}
\Tr{\Delta}_{\mu\nu}(x_{12})=\langle j_\mu(x_1)j_\nu(x_2)\rangle^{\rm tr}
\sim (\partial_\mu\partial_\nu-\delta_{\mu\nu}\Box)\Box^{\frac{D-4}{2}}
\delta(x_{12}).
\end{equation}

Another aspect of this problem consists in the fact that the conformal 
propagator of the field $A_\mu(x)$, defined by the equation~\ref{2.4} for 
$l=l_A=1$, turns out to be longitudinal
\begin{equation}\label{3.8}
\Long{D}_{\mu\nu}(x_{12})=\langle A_\mu(x_1)A_\nu(x_2)\rangle^{\rm long}
\sim \frac{1}{x_{12}^2}g_{\mu\nu}(x_{12})=\frac{1}{2}\partial_\mu\partial_\nu
\ln x_{12}^2.
\end{equation}
Note that the propagator of the regularized field $A_\mu^{D-l}$ has a 
transversal component of the order $\epsilon$
\begin{equation}\label{3.9}
\left.D_{\mu\nu}^\epsilon(x_{12})\right|_{\epsilon\to0}=
\left.\langle A_\mu^{D-l}(x_1)A_\nu^{D-l}(x_2)\rangle\right|_{\epsilon\to0}
=\Long{D}_{\mu\nu}(x_{12})+\epsilon\Tr{D}_{\mu\nu}(x_{12})+\ldots,
\end{equation}
where $\Tr{D}_{\mu\nu}(x_{12})\sim (\delta_{\mu\nu}\Box-
\partial_\mu\partial_\nu)\ln x^2$. Comparing that with the leading terms of 
the expansion~\ref{2.17}
\begin{equation}\label{3.10}
\left.\Delta_{\mu\nu}^\epsilon(x_{12})\right|_{\epsilon\to0}\simeq
\frac{1}{\epsilon}\Tr{\Delta}_{\mu\nu}(x_{12})+\Long{\Delta}_{\mu\nu}
(x_{12})+\ldots,
\end{equation}
where $\Tr{\Delta}_{\mu\nu}$ is given by the expression~\ref{3.7} and
$\Long{\Delta}_{\mu\nu}$ ---  by the expression~\ref{2.16}, one may 
conclude that the resolving of the $0\times\infty$ ambiguity should occur
in the transversal sector of the contraction~\ref{3.5}. This elucidates the 
cause of famous difficulties in attempts to analyze conformal gauge 
theories: the propagator of the gauge field, formally defined at 
$\epsilon=0$, looks like longitudinal, and the theory seems to be empty.

To inspect the problem, consider an undecomposable representation
$Q_A$ defined by the transformation~\ref{3.6} on the space $M_A$ of all 
fields $A_\mu(x)$. The space $M_A$ has an invariant subspace $\Long{M}_A$
$$\Long{M}_A \subset M_A$$
consisting from longitudinal fields
$$\Long{A}_\mu\subset \Long{M}_A,\quad
\Long{A}_\mu(x)=\partial_\mu \varphi(x).$$
Indeed, the result of conformal transformation of a longitudinal field is 
also a longitudinal field:
$$\Long{A}_\mu(x)\inv \frac{1}{x^2}g_{\mu\nu}(x)\Long{A}_\nu(Rx)=
\frac{1}{x^2}g_{\mu\nu}(x)\partial_\nu^{Rx}\varphi(Rx)=
\partial_\mu^x\varphi(Rx),$$
i.e., 
$$\Long{A}_\mu(x)\inv \Long{{A'}}_\mu(x)=\partial_\mu\varphi'(x)
\in \Long{M}_A,$$
where $\varphi'(x)=\varphi(Rx)$. It is essential that the complement of the 
subspace $\Long{M}_A$ upto the total space $M_A$ is not invariant with 
respect to transformation~\ref{3.6}. In particular, the subspace of 
transversal field is non-invariant:
$$\Tr{A}_\mu(x)\inv \tilde{A}_\mu(x)=\frac{1}{x^2}
g_{\mu\nu}(x)\Tr{A}_\nu(Rx),\quad \partial_\mu\tilde{A}_\mu(x)\ne0.$$
This feature is characteristic for any undecomposable representations.

In the case of the current $j_\mu(x)$ the undecomposable representation
is given by the transformation law
\begin{equation}\label{3.11}
j_\mu(x)\inv j_\mu'(x) = U_R^j j_\mu(x)=\frac{1}{(x^2)^{D-1}}
g_{\mu\nu}(x)j_\nu(Rx).
\end{equation}
Denote the space of representation $Q_j$ as $M_j$. It consists of all fields 
$j_\mu(x)$. Transversal fields compose an invariant subspace $\Tr{M}_j$:
$$\Tr{j}_\mu \in \Tr{M}_j \subset M_j.$$
Indeed, after the transformation of a transversal field $\Tr{j}_\mu(x)$ a new 
transversal field results:
$$\Tr{j'}_\mu(x)=\frac{1}{(x^2)^{D-1}}g_{\mu\nu}(x)\Tr{j}_\nu(Rx);\quad
\partial_\mu\Tr{j'}_\mu(x)=0\ \ \hbox{if $\partial_\mu\Tr{j}_\mu(x)=0$.}$$
To check that, the relation 
$\partial_\mu\left[\frac{1}{(x^2)^{D-1}}g_{\mu\nu}(x)\right]=0$ is used.
A complement of $\Tr{M}_j$ to the total space $M_j$ is non-invariant.
Thus the transformation law~\ref{3.11} does not define a representation on 
the space of longitudinal currents $\Long{j}_\mu(x)=\partial_\mu j(x)$.

Now consider invariant bilinear forms for even $D\ge4$. On the space $M_A$ 
exists a unique finite form
\begin{equation}\label{3.12}
\left\{A,A\right\}_0 = 
\int dx_1\,dx_2\,
A_\mu(x_1)\Tr{\Delta}_{\mu\nu}(x_{12})A_\nu(x_2),
\end{equation}
where $\Tr{\Delta}_{\mu\nu}$ is the exceptional kernel~\ref{3.7}. This form 
is degenerate on the invariant subspace $\Long{M}_A$
\begin{equation}\label{3.13}
\left\{\Long{A},\Long{A}\right\}_0=0.
\end{equation}
Analogously, the only invariant form on the space $M_j$ has the form
\begin{equation}\label{3.14}
\left\{j,j\right\}_0=
\int dx_1\,dx_2\,
j_\mu(x_1)\Long{D}_{\mu\nu}(x_{12})j_\nu(x_2).
\end{equation}
It is degenerate on the invariant subspace $\Tr{M}_j$:
\begin{equation}\label{3.15}
\left\{\Tr{j},\Tr{j}\right\}_0=0.
\end{equation}

To obtain non-degenerate forms one should consider irreducible 
representations instead of undecomposable representations $Q_A$ and $Q_j$.
To each undecomposable representation $Q$ defined on the space $M$ one can 
attach a pair of irreducible representations. One of them acts on an 
invariant subspace $M_0\subset M$, namely, on the subspace where the invariant 
form is degenerate, see \ref{3.12}--\ref{3.15}. Denote this 
representation as 
$Q_0$. It is defined by the initial transformation law. The other irreducible 
representation is established by the same transformation law on the quotient 
space
$$\tilde{M}=M/M_0.$$
Denote this representation as $\tilde{Q}$. The elements of the space are 
equivalence classes.

In the case of the potential $A_\mu$ the quotient space 
$$\tilde{M}_A=M/\Long{M}$$
consists of equivalence classes $\left[A_\mu\right]$, each class including 
all fields with a fixed transversal component. Any two fields $A_\mu$
and $A_\mu+\partial_\mu\varphi$, where $\varphi$ is an arbitrary function, 
are the representatives of the same class $\left[A\right]$. The 
transformation law~\ref{3.6} defines an irreducible representation on such 
classes:
\begin{equation}\label{3.16}
\tilde{Q}_A:\quad \left[A_\mu\right]\inv\left[A_\mu'\right].
\end{equation}    
The form~\ref{3.12} gives the scalar product on these classes. 
One may choose transversal representatives in each class --- it does not
effect the value of the form:  
\begin{equation}\label{3.17} 
\int dx_1\,dx_2\, A_\mu(x_1)\Tr{\Delta}_{\mu\nu}(x_{12})A_\nu(x_2)= \int 
dx_1\,dx_2\, \Tr{A}_\mu(x_1)\Tr{\Delta}_{\mu\nu}(x_{12}) \Tr{A}_\nu(x_2).  
\end{equation}
It follows that the form $\left\{A,A\right\}_0$ is non-degenerate when 
exploited on the quotient space $\tilde{M}_A$. The other 
irreducible representation, of the $Q_0$ type, is given by the transformation
law~\ref{3.6} on an invariant subspace $\Long{M}_A$:
$$\Long{A}_\mu(x)\inv \Long{A'}_\mu(x).$$
Denote this representation as $\Long{Q}_A$, and consider its invariant scalar 
product. The latter may be defined by the form
\begin{equation}\label{3.18}
\left\{\Long{A},\Long{A}\right\}_1 = 
\int dx_1\,dx_2\, 
\Long{A}_\mu(x_1)\Long{\Delta}_{\mu\nu}(x_{12})\Long{A}_\nu(x_2),
\end{equation} 
where $\Long{\Delta}_{\mu\nu}$ is the longitudinal kernel~\ref{2.16}.
This form is invariant only on the subspace of longitudinal fields. 
Its expression may be chosen to be explicitly invariant 
using the singular kernel~\ref{2.17}:
\begin{equation}\label{3.19}
\int dx_1\,dx_2\, 
\Long{A}_\mu(x_1)\Long{\Delta}_{\mu\nu}(x_{12})\Long{A}_\nu(x_2)=
\left.\int dx_1\,dx_2\, 
\Long{A}_\mu(x_1)\Delta^\epsilon_{\mu\nu}(x_{12})
\Long{A}_\nu(x_2)\right|_{\epsilon=0}.
\end{equation}

Thus we get the pair of irreducible representations
\begin{equation}\label{3.20}
\tilde{Q}_A\quad\hbox{and}\quad\Long{Q}_A
\end{equation}
acting on the spaces $\tilde{M}_A$ and $\Long{M}_A$. Invariant scalar 
products are defined on both of these spaces. Accordingly, there is the
pair of invariant kernels
\begin{equation}\label{3.21}
\Tr{\Delta}_{\mu\nu}(x_{12})\quad\hbox{and}\quad
\Delta^\epsilon_{\mu\nu}(x_{12}).
\end{equation}
The former one is non-degenerate on the space $\tilde{M}_A$, while the
latter is non-degenerate (and finite) on the space $\Long{M}_A$.
Now let us consider a direct sum of irreducible representations
\begin{equation}\label{3.22}
\tilde{Q}_A \oplus \Long{Q}_A,
\end{equation}
defined on the direct sum of spaces
\begin{equation}\label{3.23}
\tilde{M}_A \oplus \Long{M}_A.
\end{equation}
As will be shown in the next section, having chosen a suitable realization of 
conformal transformations one would be able to construct an invariant scalar 
product on the space~\ref{3.23}
\begin{equation}\label{3.24}
(A,A)=\left\{\Tr{A},\Tr{A}\right\}_0+
\left\{\Long{A},\Long{A}\right\}_1,
\end{equation}
which would define a new propagator of the current.

In an analogous manner one can explore the pair of irreducible 
representations
\begin{equation}\label{3.25}
\Tr{Q}_j\quad\hbox{and}\quad\tilde{Q}_j,
\end{equation}
associated with an undecomposable representation $Q_j$. The first one,
$\Tr{Q}_j$, is given by the transformation law~\ref{3.11} on the invariant 
space $\Tr{M}_j$:
\begin{equation}\label{3.26}
\Tr{Q}_j:\quad \Tr{j}_\mu(x)\inv \Tr{j'}_\mu.
\end{equation}
The second irreducible representation is given by the law~\ref{3.11} on the 
quotient space $\tilde{M}_j=M_j/\Tr{M}_j$ which consists of equivalence 
classes $\left[j_\mu\right]$:
\begin{equation}\label{3.27}
\tilde{Q}_j:\quad \left[j_\mu\right]\inv \left[j_\mu'\right].
\end{equation}
Each class contains all fields with a fixed longitudinal component. Any two 
fields $j_\mu$ and $j_\mu+\Tr{\tilde{j}}_\mu$, where $\Tr{\tilde{j}}_\mu$ is 
a transversal field, belong to the same class. The invariant form~\ref{3.14} 
defines a scalar product on these equivalence classes. In what follows in 
each class we select a longitudinal representative. The value of the form 
will be the same after that:
\begin{equation}\label{3.28}
\int dx_1\,dx_2\,
j_\mu(x_1)\Long{D}_{\mu\nu}(x_{12})j_\nu(x_2) 
= \int dx_1\,dx_2\,
\Long{j}_\mu(x_1)\Long{D}_{\mu\nu}(x_{12})\Long{j}_\nu(x_2). 
\end{equation}
Consider the invariant scalar product for the representation $\Tr{Q}_j$. The 
latter may be defined by the form
\begin{equation}\label{3.29}
\left\{\Tr{j},\Tr{j}\right\}_1=\int dx_1\,dx_2\,
\Tr{j}_\mu(x_1)\Tr{D}_{\mu\nu}(x_{12})\Tr{j}_\nu(x_2),
\end{equation}
where
\begin{equation}\label{3.30}
\Tr{D}_{\mu\nu}\sim \left(\delta_{\mu\nu}\Box-\partial_\mu\partial_\nu\right)
\ln x_{12}^2.
\end{equation}
Though the kernel $\Tr{D}_{\mu\nu}$ is not invariant under conformal 
inversion
\begin{eqnarray*} 
&& \Tr{D}_{\mu\nu}(x_{12})=\frac{1}{x_1^2x_2^2}
g_{\mu\rho}(x_1)g_{\nu\sigma}(x_2)\Tr{D}_{\mu\nu}(Rx_1-Rx_2) \\
&&{}-\frac{1}{x_1^2}g_{\mu\tau}(x_1)\partial_\tau^{Rx_1}
\left(\frac{1}{2}\ln x_{12}^2\partial_\nu\ln x_2^2\right)
-\frac{1}{x_2^2}g_{\mu\tau}(x_2)\partial_\tau^{Rx_2}
\left(\frac{1}{2}\ln x_{12}^2\partial_\nu\ln x_1^2\right)\\
&&\quad\quad{}-\frac{1}{x_1^2x_2^2}g_{\mu\tau}(x_1)g_{\nu\sigma}(x_2)
\partial_\tau^{Rx_1}\partial_\sigma^{Rx_2}\left(\ln x_1^2
\ln x_2^2\right),
\end{eqnarray*}
the form~\ref{3.29} remains invariant as long as considered
on transversal fields. Note that the form~\ref{3.29} may be written as an 
explicitly invariant expression if one introduces the invariant kernel
which is singular when $\epsilon\to0$:
\begin{equation}\label{3.31}
D_{\mu\nu}^\epsilon(x_{12})\sim\frac{1}{\epsilon}
\frac{1}{(x_{12})^{1-\epsilon}}g_{\mu\nu}(x_{12})=
\frac{1}{\epsilon}\Long{D}_{\mu\nu}(x_{12})+
\Tr{D}_{\mu\nu}(x_{12})+\ldots.
\end{equation}
The contractions of this kernel with transversal fields are finite in the 
limit $\epsilon=0$. Now rewrite~\ref{3.29} in the form
\begin{equation}\label{3.32}
\left\{\Tr{j},\Tr{j}\right\}_1=\left.\int dx_1\,dx_2\,
\Tr{j}_\mu(x_1)D_{\mu\nu}^\epsilon(x_{12})\Tr{j}_\nu(x_2)\right|_{\epsilon=0}.
\end{equation}
Here we assume the fields $\Tr{j}_\mu$ are subjected to conformally invariant
regularization (see sections~4,5):
$$\left.\partial_\mu\Tr{j}_\mu(x)\right|_{\epsilon=0}=0.$$

Resultantly, we have a pair of invariant kernels (in accordance 
with~\ref{3.21})
\begin{equation}\label{3.33}
D_{\mu\nu}^\epsilon(x_{12})\quad\hbox{and}\quad\Long{D}_{\mu\nu}(x_{12}).
\end{equation}
The first kernel is non-degenerate and finite on the space $\tilde{M}_A$,
and the second --- on the subspace $\Long{M}_A$. Analogously to~\ref{3.22},
consider a direct sum of representations~\ref{3.20}:
\begin{equation}\label{3.34}
\Tr{Q}_j \oplus \tilde{Q}_j,
\end{equation}
defined on the direct sum of spaces
\begin{equation}\label{3.35}
\Tr{M}_j \oplus \tilde{M}_j.
\end{equation}
In the next section we derive a non-degenerate propagator $\langle A_\mu
A_\nu\rangle$ defining the latter as a kernel of the invariant product
\begin{equation}\label{3.36}
(j,j)=\left\{\Tr{j},\Tr{j}\right\}_1+
\left\{\Long{j},\Long{j}\right\}_0=
\int dx_1\,dx_2\, j_\mu(x_1)\langle A_\mu(x_1) A_\nu(x_2)\rangle
j_\nu(x_2)
\end{equation}
in a specific realization of conformal transformations.

\subsection{Irreducible Representations for the Fields
$h_{\mu\nu}$ and $T_{\mu\nu}$} 
Define a propagator of the energy-momentum tensor 
\begin{equation}\label{3.37}
\Delta_{\mu\nu\rho\sigma}(x_{12})=
\langle T_{\mu\nu}(x_1)T_{\rho\sigma}(x_2)\rangle
\end{equation}
as the kernel of the invariant form
\begin{equation}\label{3.38}
(h,h)=\int dx_1\,dx_2\,
h_{\mu\nu}(x_1)\langle T_{\mu\nu}(x_1)T_{\rho\sigma}(x_2)\rangle
h_{\rho\sigma}(x_2),
\end{equation}
where $h_{\mu\nu}$ is the traceless part of the metric tensor. Consider the 
transformation law
\begin{equation}\label{3.39}
h_{\mu\nu}(x)\inv h_{\mu\nu}'(x)=U_R^h h_{\mu\nu}(x)=
g_{\mu\rho}(x)g_{\nu\sigma}(x)h_{\rho\sigma}(Rx).
\end{equation}
If the form~\ref{3.38} were invariant with respect to this law, the 
propagator~\ref{3.37} would satisfy the equation~\ref{2.19} for $l=D$.
We know that the solution of the equation~\ref{2.19} does not allow a 
limiting transition $l\to D$ for even $D\ge4$. Nevertheless, there exists
an exceptional solution
\begin{equation}\label{3.40}
\Tr{\Delta}_{\mu\nu\rho\sigma}(x_{12}) = 
\langle T_{\mu\nu}(x_1)T_{\rho}(x_2)\rangle^{\rm tr}\sim
\Tr{H}_{\mu\nu\rho\sigma}\left(\frac{\partial}{\partial x}\right)
\Box^{\frac{D-4}{2}}\delta(x_{12}),
\end{equation}
where $\Tr{H}_{\mu\nu\rho\sigma}$ is the differential operator~\ref{224}.
This solution cannot be derived from the general solution~\ref{2.20} as the
limit $l\to D$. Therefore, the propagator of the field $T_{\mu\nu}(x)$,
which is invariant under the transformation
\begin{equation}\label{3.41}
T_{\mu\nu}(x)\inv T_{\mu\nu}'(x)=U_R^T T_{\mu\nu}(x)=
\frac{1}{(x^2)^D}g_{\mu\rho}(x)g_{\nu\sigma}(x)T_{\rho\sigma}(Rx),
\end{equation}
is transversal. On the other hand, the propagator of the field 
$h_{\mu\nu}(x)$, being invariant under~\ref{3.39}, is longitudinal. The 
solution of the equation~\ref{2.19} for $l=0$ reads:
\begin{eqnarray} \label{3.42}
&& \Long{D}_{\mu\nu\rho\sigma}(x_{12})=
\langle h_{\mu\nu}(x_1)h_{\rho\sigma}(x_2)\rangle^{\rm long}\nonumber\\
&&\qquad\qquad{}\sim\left[g_{\mu\rho}(x_{12})
g_{\nu\sigma}(x_{12})+g_{\mu\sigma}(x_{12})g_{\nu\rho}(x_{12})-
\frac{2}{D}\delta_{\mu\nu}\delta_{\rho\sigma}\right]\nonumber\\
&&\quad\quad{} = \partial_\mu D_{\nu\rho\sigma}(x_{12})+
\partial_\nu D_{\mu\rho\sigma}(x_{12})-\frac{2}{D}\delta_{\mu\nu}
\partial_\lambda D_{\lambda\rho\sigma}(x_{12}),
\end{eqnarray}
where 
$$D_{\mu,\rho\sigma}(x) = \frac{1}{2}\left[
x_\rho g_{\nu\sigma}(x)+x_\sigma g_{\mu\rho}(x)+
\frac{2}{D}\delta_{\rho\sigma} x_\mu\right].$$
Hence it follows that similar to the case of current 
(see~\ref{3.9},\ref{3.10} and below), the transversal sector of the 
contraction~\ref{3.38} exhibits an ambiguity $0\times\infty$.

An inspection of undecomposable representations $Q_h$ and $Q_T$ defined by 
the transformation laws~\ref{3.39} and~\ref{3.41} is a literal recital of the 
analysis of representations $Q_A$ and $Q_j$ studied above. So, we shall 
restrict ourselves to a brief discourse. Look at the representation
$Q_h$ defined by the law~\ref{3.39} in the space $M_h$ of fields 
$h_{\mu\nu}$. This space has an invariant subspace $\Long{M}_h$, consisting
of longitudinal fields
\begin{equation}\label{3.43}
\Long{h}_{\mu\nu}(x) = \partial_\mu h_\nu(x)+
\partial_\nu h_\mu(x)-\frac{2}{D}\delta_{\mu\nu}\partial_\lambda 
h_\lambda(x).  
\end{equation} 
The transformation law~\ref{3.39} defines the irreducible representation
$\Long{Q}_h$ in the above space. The form~\ref{3.38} with the 
kernel~\ref{3.40} vanishes on $\Long{M}_h$. An invariant scalar product may 
be defined by the form~\ref{2.24}, the latter will be denoted as
$\left\{\Long{h},\Long{h}\right\}_1$. The second irreducible representation,
denoted as $\tilde{Q}_h$, is given by the law~\ref{3.39} in the quotient 
space
$$\tilde{M}_h=M_h/\Long{M}_h.$$
The elements of this space are equivalence classes, each including all fields
with a fixed transversal component. Two fields $h_{\mu\nu}$ and
$h_{\mu\nu}+\Long{h}_{\mu\nu}$ belong to the same class. The invariant scalar
product on $\tilde{M}_h$ is given by the form
\begin{equation}\label{3.44}
\left\{h,h\right\}_0=\int dx_1\,dx_2\,
h_{\mu\nu}(x_1)\Tr{\Delta}_{\mu\nu\rho\sigma}(x_{12})
h_{\rho\sigma}(x_2),
\end{equation}
where $\Tr{\Delta}_{\mu\nu\rho\sigma}$ is the transversal kernel~\ref{3.40}.
Thus one has a pair of irreducible representations
\begin{equation}\label{3.45}
\tilde{Q}_h\quad\hbox{and}\quad\Long{Q}_h,
\end{equation}
associated with the pair of invariant kernels
\begin{equation}\label{3.46}
\Tr{\Delta}_{\mu\nu\rho\sigma}(x_{12})\quad\hbox{and}\quad
\Delta^\epsilon_{\mu\nu\rho\sigma}(x_{12}).
\end{equation}
The second kernel is given by the expression~\ref{2.30} formally divergent at
$\epsilon=0$, but gives rise to the form $\left\{,\right\}_1$ which is finite 
on $\Long{M}_h$. Examine a direct sum of representations
\begin{equation}\label{3.47}
\tilde{Q}\oplus\Long{Q}_h.
\end{equation}
An invariant scalar product may be postulated as a sum of forms
\begin{equation}\label{3.48}
(h,h)=\left\{\Tr{h},\Tr{h}\right\}_0+
\left\{\Long{h},\Long{h}\right\}_1.
\end{equation}
In the next section we study a realization of representation $\tilde{Q}$
on the space of transversal fields $\Tr{h}_{\mu\nu}$ and obtain the 
expression for the propagator $\langle T_{\mu\nu}T_{\rho\sigma}\rangle$,
given by the formula~\ref{3.38}.

An undecomposable representation $Q_T$ given by the law~\ref{3.41} may be
inspected in a similar manner. The space $M_T$ of the representation
$Q_T$ has an invariant subspace
$$\Tr{M}_T \subset M_T.$$
Indeed, the transformation of a transversal field $\Tr{T}_{\mu\nu}$ results
in a transversal field $\Tr{T'}_{\mu\nu}$:
$$\partial_\mu \Tr{T'}_{\mu\nu}(x)=
\partial_\mu\left\{\frac{1}{(x^2)^D}g_{\mu\rho}(x)g_{\nu\sigma}(x)
T_{\rho\sigma}(Rx)\right\}=0.$$
To check that, one uses the equality 
$\partial_\mu\left[\frac{1}{(x^2)^{D-1}}g_{\mu\rho}(x)\right]=0$
and the fact that the tensor $T_{\mu\nu}(x)$ is traceless. The transformation
law~\ref{3.41} defines an irreducible representation $\Tr{Q}_T$ on the
space $\Tr{M}_T$. An invariant scalar product on that space may be defined
by the form
\begin{equation}\label{3.49}
\left\{\Tr{T},\Tr{T}\right\}_1=\left.\int dx_1\,dx_2\,
\Tr{T}_{\mu\nu}(x_1)D_{\mu\nu\rho\sigma}^\epsilon(x_{12})
\Tr{T}_{\rho\sigma}(x_2)\right|_{\epsilon=0},
\end{equation}
where
\begin{equation}\label{3.50}
D_{\mu\nu\rho\sigma}(x)\sim\frac{1}{\epsilon}(x^2)^\epsilon\left[
g_{\mu\rho}(x)g_{\nu\sigma}(x)+g_{\mu\sigma}(x)
g_{\nu\rho}(x)- \frac{2}{D}\delta_{\mu\nu}\delta_{\rho\sigma}\right].
\end{equation}
Though this kernel is formally divergent, the contractions~\ref{3.49} are
always finite due to~\ref{3.42}. A conformally invariant regularization
of transversal fields $\Tr{T}_{\mu\nu}(x)$ in~\ref{3.49} is detailed in
section~5. Note that the form~\ref{3.49} can be defined using the formally
non-invariant kernel $\Tr{D}_{\mu\nu\rho\sigma}$ (similar to~\ref{3.29}):
\begin{equation}\label{3.51}
\left.\int dx_1\,dx_2\,
\Tr{T}_{\mu\nu}(x_1)D_{\mu\nu\rho\sigma}^\epsilon(x_{12})
\Tr{T}_{\rho\sigma}(x_2)\right|_{\epsilon=0} = 
\left.\int dx_1\,dx_2\,
\Tr{T}_{\mu\nu}(x_1)\Tr{D}_{\mu\nu\rho\sigma}(x_{12})
\Tr{T}_{\rho\sigma}(x_2)\right|_{\epsilon=0},
\end{equation}
where
\begin{equation}\label{3.52}
\Tr{D}_{\mu\nu\rho\sigma}(x)\sim\Tr{P}_{\mu\nu\rho\sigma}
\left(\frac{\partial}{\partial x}\right)
\ln x^2,
\end{equation}
and $P\left(\frac{\partial}{\partial x}\right)$ is the projection
operator~\ref{2.35}. The second irreducible representation, denoted as
$\tilde{Q}_T$, acts in the quotient space (similar to~\ref{3.27})
$$\tilde{M}_T=M_T/\Tr{M}_T.$$
The invariant scalar product on $\tilde{M}_T$ is defined by the form
\begin{equation}\label{3.53}
\left\{T,T\right\}_0 = \int dx_1\,dx_2\,
T_{\mu\nu}(x_1)\Long{D}_{\mu\nu\rho\sigma}(x_{12})T_{\rho\sigma}(x_2).
\end{equation}
Choosing a longitudinal representative in each equivalence class 
$\tilde{M}_T$
\begin{equation}\label{354}
\Long{T}_{\mu\nu}(x)=
\partial_\mu T_\nu(x)+\partial_\nu T_\mu(x)-\frac{2}{D}\delta_{\mu\nu}
\partial_\lambda T_\lambda(x),
\end{equation}
one can rewrite~\ref{3.53} in the form (see~\ref{3.28}): 
\begin{equation}\label{355}
\left\{T,T\right\}_0=\int dx_1\,dx_2\,
\Long{T}_{\mu\nu}(x_1)\Long{D}_{\mu\nu\rho\sigma}(x_{12})
\Long{T}_{\rho\sigma}(x_2).
\end{equation}
Thus one has a pair of irreducible representations
\begin{equation}\label{3.54}
\Tr{Q}_T\quad\hbox{and}\quad\tilde{Q}_T,
\end{equation}
associated with the pair of invariant forms
\begin{equation}\label{3.55}
\left\{\Tr{T},\Tr{T}\right\}_1\quad\hbox{and}\quad
\left\{\Long{T},\Long{T}\right\}_0.
\end{equation}
In the next section we examine the field $T_{\mu\nu}(x)$ which transforms
by the irreducible representation
\begin{equation}\label{3.56}
\Tr{Q}_T \oplus \tilde{Q}_T.
\end{equation}
Choosing a special realization of the representation $\tilde{Q}_T$,
we shall introduce an invariant propagator $\langle h_{\mu\nu}
h_{\rho\sigma}\rangle$ as the kernel of the form
\begin{equation}\label{3.57}
(T,T)=\left\{\Tr{T},\Tr{T}\right\}_1+\left\{\Long{T},\Long{T}\right\}_2
=\int dx_1\,dx_2\,
T_{\mu\nu}(x_1)\langle h_{\mu\nu}(x_1)h_{\rho\sigma}(x_2)\rangle
T_{\rho\sigma}(x_2).
\end{equation}

\section{New Conformal Transformations and Conformal Propagators
$\langle j_\mu j_\nu\rangle,\langle T_{\mu\nu}T_{\rho\sigma}\rangle$}
\setcounter{equation}{0}
Consider the irreducible representation $\tilde{Q}_A$, see~\ref{3.16}.
The transformation law~\ref{3.6} defines it on the space $\tilde{M}_A$
of equivalence classes. Let us present a new realization of this 
representation on the space of transversal fields. Of course, the 
transformation law of the field $A_\mu(x)$ will be different in this
realization.

First, in each equivalence class $\left[A_\mu\right]\subset\tilde{M}_A$ we 
pass to a transversal representative $\Tr{A}_\mu\subset\left[A_\mu\right]$:
\begin{eqnarray} \label{4.1}
&& A_\mu(x)\to \Tr{A}_\mu(x) = \Tr{P}A_\mu(x)=
\left(\delta_{\mu\nu}-\frac{\partial_\mu\partial_\nu}{\Box}\right)
A_\nu(x),\nonumber \\
&&A_\mu,\Tr{A}_\mu(x) \subset\left[A_\mu\right].
\end{eqnarray}
The transformation~\ref{3.6} converts the class $\left[A_\mu\right]$
into a new class $\left[A_\mu'\right]$, see~\ref{3.16}. Under that, a
transversal representative $\Tr{A}_\mu$ transforms into a certain
(non-transversal) field $A_\mu'\subset\left[A_\mu'\right]$
\begin{equation}\label{4.2}
A_\mu'(x) = U_R^A \Tr{A}_\mu(x) = \frac{1}{x^2}g_{\mu\nu}(x)
\Tr{A}_\nu(Rx).
\end{equation} 
In the new class $\left[A_\mu'\right]$ we pass to a transversal 
representative $\Tr{A'}_\mu$
\begin{equation}\label{4.3}
\Tr{A'}_\mu(x) = \Tr{P} A_\mu'(x) = 
\left(\delta_{\mu\nu}-\frac{\partial_\mu\partial_\nu}{\Box}\right)
A_\nu'(x).
\end{equation}
The sequence of three transformation~\ref{4.1}--\ref{4.3} defines a 
realization of the representation $\tilde{Q}_A$ on transversal fields.
Taking into account that longitudinal fields $\Long{A}_\mu$ transform
by the irreducible representation $\Long{Q}_A$, one can now define a 
reducible representation $\tilde{Q}_A\oplus\Long{Q}_A$ on the total space
of fields $A_\mu(x)$ as follows. Decompose the field $A_\mu$ into the
sum of transversal and longitudinal components
$$A_\mu(x) = \Tr{A}_\mu(x)+\Long{A}_\mu(x) = 
\Tr{P} A_\mu(x)+\Long{P}A_\mu(x).$$
With the first term we associate the representation $\tilde{Q}_A$, 
and with the second --- $\Long{Q}_A$. To proceed, introduce the operator
\begin{equation}\label{4.4}
V_R^A = \Tr{P} U_R^A \Tr{P}+U_R^A\Long{P},
\end{equation}
where $U_R^A$ is given by~\ref{3.6}. The transformation law
\begin{equation}\label{4.5}
A_\mu(x)\inv A_\mu'(x)=V_R A_\mu(x)
\end{equation}
defines the reducible representation $\tilde{Q}_A\oplus\Long{Q}_A$.
An expanded form of the transformation~\ref{4.5} is:
\begin{eqnarray} \label{4.6}
&& V_R A_\mu(x) = 
\left(\delta_{\mu\nu}-\frac{\partial_\mu^x\partial_\nu^x}{\Box_x}\right)
\frac{1}{x^2}g_{\nu\rho}(x)
\left(\delta_{\rho\sigma}-\frac{\partial_\rho^{Rx}
\partial_\sigma^{Rx}}{\Box_{Rx}}\right)
A_\sigma(Rx)\nonumber \\
&&\qquad\qquad\qquad{}+\frac{1}{x^2}g_{\mu\nu}(x)
\frac{\partial_\nu^{Rx}\partial_\rho^{Rx}}{\Box_{Rx}}
A_\rho(Rx)\nonumber\\
&&{}=\frac{1}{x^2}g_{\mu\nu}(x) A_\nu(Rx)-
\frac{\partial_\mu^x\partial_\nu^x}{\Box_x}
\frac{1}{x^2}g_{\nu\rho}(x)A_\rho(Rx)+\frac{1}{x^2}g_{\mu\nu}(x)
\frac{\partial_\nu^{Rx}\partial_\rho^{Rx}}{\Box_{Rx}}
A_\rho(Rx). \nonumber \\
\end{eqnarray}
Evidently, each conformal transformation $g:\ x_\mu\to gx_\mu$ of the
field $A_\mu(x)$ in the new realization may be put into a form~\ref{4.4}
\begin{equation}\label{4.7}
A_\mu(x)\stackrel{g}{\longrightarrow}V_g A_\mu(x) = 
\Tr{P}U_g\Tr{P}A_\mu(x)+U_g\Long{P}A_\mu(x),
\end{equation}
where $U_g$ is the transformation operator of the field $A_\mu(x)$
in a conventional realization. One easily checks that the 
transformation~\ref{4.7} satisfies the group law:
\begin{equation}\label{4.8}
V_{g_2}V_{g_1}A_\mu(x)=V_{g_1g_2}A_\mu(x),\quad
\hbox{if $U_{g_2}U_{g_1}A_\mu(x)=U_{g_1g_2}A_\mu(x).$}
\end{equation}
The check employs an invariance of longitudinal sector (the subspace
$\Long{M}_A$) with respect to transformations $U_g$ in conventional
realization: $\Tr{P}U_g\Long{P}A_\mu(x)=0$.

The realization of representation $\Tr{Q}_j\oplus\tilde{Q}_j$, see
\ref{3.33}--\ref{3.36}, on the space of fields $j_\mu(x)$ may be 
deduced in an analogous manner. It amounts to introducing of the 
representation $\tilde{Q}_j$, see~\ref{3.27} and below, on the space
of fields $\Long{j}_\mu(x)$. As above, we choose a longitudinal 
representative in each equivalence class $\left[j_\mu\right]$
and consider a sequence of transformations like~\ref{4.1}--\ref{4.3}
\begin{eqnarray}\label{4.9}
&& j_\mu(x)\to
\frac{\partial_\mu\partial_\nu}{\Box}
j_\nu(x) \inv \frac{1}{(x^2)^{D-1}} g_{\mu\nu}(x)
\frac{\partial_\nu^{Rx}\partial_\rho^{Rx}}{\Box_{Rx}}
j_\rho(Rx) \nonumber \\ 
&&\qquad\quad{}\to
\frac{\partial_\mu^x\partial_\nu^x}{\Box_x}\frac{1}{(x^2)^{D-1}}
g_{\nu\rho}(x)
\frac{\partial_\rho^{Rx}\partial_\sigma^{Rx}}{\Box_{Rx}}j_\sigma(Rx).
\end{eqnarray}
Using the invariance of transversal sector (the subspace $\Tr{M}_j$)
of the fields $j_\mu(x)$ under the transformation~\ref{3.11}
\begin{equation}\label{4.10}
\Long{P}U_R^j\Tr{P}j_\mu(x)=0,
\end{equation} 
we arrive at the following realization of the representation
$\Tr{Q}_j\oplus\tilde{Q}$:
\begin{equation}\label{4.11}
j_\mu(x)\inv V_R^j j_\mu(x)=
U_R^j\Tr{P}j_\mu(x)+
\Long{P}U_R^j\Long{P}j_\mu(x),
\end{equation}
or, more explicitly
\begin{eqnarray*}
&& V_R^j j_\mu(x) =
\frac{1}{(x^2)^{D-1}} g_{\mu\nu}(x)
\left(\delta_{\nu\rho}-
\frac{\partial_\nu^{Rx}\partial_\rho^{Rx}}{\Box_{Rx}}\right)
j_\rho(Rx)\nonumber\\
&&\qquad\qquad{}+
\frac{\partial_\mu^x\partial_\nu^x}{\Box_x}
\frac{1}{(x^2)^{D-1}} g_{\nu\rho}(x)
\frac{\partial_\rho^{Rx}\partial_\sigma^{Rx}}{\Box_{Rx}} j_\sigma(Rx).
\end{eqnarray*}
Here the first term corresponds to the representation $\Tr{Q}_j$, while
the second --- to the representation $\tilde{Q}_j$. The group law
is checked in the same way as in the case of the field $A_\mu(x)$,
see~\ref{4.8}. On account of~\ref{4.10}, we have:
$$V_R^j V_R^j j_\mu(x)=j_\mu(x),\quad\hbox{since}\quad
U_R^j U_R^j j_\mu(x)=j_\mu(x).$$

Thus, we have formulated a new realization of conformal transformations
of the fields $A_\mu$ and $j_\mu$. Now consider invariant (in the new sense)
propagators of these fields. The conditions of invariance have the form
\begin{eqnarray} 
\label{4.12} 
\langle V_R^A A_\mu(x_1)V_R^A A_\nu(x_2)\rangle  & =  &
\langle A_\mu(x_1)A_\nu(x_2)\rangle, \\
\label{4.13}
\langle V_R^j j_\mu(x_1)V_R^j j_\nu(x_2)\rangle  & =  &
\langle j_\mu(x_1)j_\nu(x_2)\rangle.
\end{eqnarray}
The solution of these equations has the form~[1,14--16]:
\begin{eqnarray} 
\label{4.14}
D_{\mu\nu}(x_{12}) = \langle A_\mu(x_1)A_\nu(x_2) \rangle & = &
g_A \left(\delta_{\mu\nu}-\frac{\partial_\mu\partial_\nu}{\Box} \right)
\frac{1}{x_{12}^2} + \eta\partial_\mu\partial_\nu\ln x_{12}, \\
\label{4.15}
\Delta_{\mu\nu}(x_{12}) = \langle j_\mu(x_1)j_\nu(x_2) \rangle & = &
g_j \left(\delta_{\mu\nu}-\frac{\partial_\mu\partial_\nu}{\Box} \right)
\Box^{\frac{D-4}{2}}\delta(x_{12}) + C_j \partial_\mu\partial_\nu
\Box^{\frac{D-4}{2}}\delta(x_{12}), \nonumber \\
\end{eqnarray}
where $g_A,\eta,g_j$ are constants. The coefficient $C_j$ in the second
term of~\ref{4.15} is chosen from the Ward identity~\ref{2.12}. The 
propagators ~\ref{4.14} and~\ref{4.15} are the kernels of conformally 
invariant (in the new sense) contractions~\ref{3.24} and~\ref{3.35}, 
see also~\ref{3.5}.

Consider the fields $h_{\mu\nu}(x)$ and $T_{\mu\nu}(x)$. Repeating
the derivation of transformations~\ref{4.4} and~\ref{4.11} literally,
we get the new conformal transformations in the following form~[1],
see also~[17]:
\begin{eqnarray} 
&& \label{4.16}
h_{\mu\nu} (x) \inv V_R^h h_{\mu\nu} (x) = 
\Tr{P}U_R^h\Tr{P}h_{\mu\nu}(x)+U_R^h \Long{P}h_{\mu\nu}(x), \\
&& \label{4.17}
T_{\mu\nu} (x) \inv V_R^T T_{\mu\nu} (x) = 
U_R^T \Tr{P}T_{\mu\nu}(x)+\Long{P}U_R^T\Long{P}T_{\mu\nu}(x),
\end{eqnarray}
where $U_R^h$ and $U_R^T$ are given by the relations~\ref{3.39} 
and~\ref{3.41}, and the projection operators $\Tr{P},\Long{P}$ are written
down in~\ref{2.35} and~\ref{2.33}. These transformations define reducible
representations of the conformal group, see~\ref{3.45}--\ref{3.48} and
\ref{3.54}--\ref{3.57}:
\begin{equation}\label{4.18}
\tilde{Q}_h\oplus\Long{Q}_h\quad\hbox{and}\quad
\Tr{Q}_T\oplus\tilde{Q}_T.
\end{equation}  
In the derivation of~\ref{4.16},\ref{4.17} we have used the invariance of 
longitudinal sector of the field $h_{\mu\nu}(x)$ under the action of
$U_R^h$, and the invariance of transversal sector of the field 
$T_{\mu\nu}(x)$ under the action of $U_R^T$ (the subspaces $\Long{M}_h$
and $\Tr{M}_T$)
$$\Tr{P}U_R^h \Long{P}h_{\mu\nu}(x) = 0,\quad
\Long{P}U_R^T \Tr{P}T_{\mu\nu}(x) = 0.$$
The conformally invariant (in the new sense) propagators of the fields
$h_{\mu\nu}(x)$ and $T_{\mu\nu}(x)$ are found from the conditions of 
invariance, analogous to~\ref{4.12},\ref{4.13}. They have the form
(see also~[1,17] and references therein):
\begin{eqnarray}
&& \label{4.19}
D_{\mu\nu\rho\sigma}(x_{12}) = 
\langle h_{\mu\nu}(x_1)h_{\rho\sigma}(x_2)\rangle\nonumber \\
&&\qquad\qquad{} = g_h \Tr{P}_{\mu\nu\rho\sigma}
\left(\frac{\partial}{\partial x}\right)
\ln x_{12}^2+\eta \Long{P}_{\mu\nu\rho\sigma}
\left(\frac{\partial}{\partial x}\right)
\ln x_{12}^2, \\
&& \label{4.20}
\Delta_{\mu\nu\rho\sigma}(x_{12}) = 
\langle T_{\mu\nu}(x_1)T_{\rho\sigma}(x_2)\rangle\nonumber \\
&&\qquad\qquad{} = g_T \Tr{H}_{\mu\nu\rho\sigma}
\left(\frac{\partial}{\partial x}\right)
\Box^{\frac{D-4}{2}}\delta(x_{12})+
C_T \Long{H}_{\mu\nu\rho\sigma}
\left(\frac{\partial}{\partial x}\right)
\Box^{\frac{D-4}{2}}\delta(x_{12}). 
\end{eqnarray}  
Here $g_h$, $\eta$ and $g_T$ are constants, the projection operators
$\Tr{P}_{\mu\nu\rho\sigma}$ and $\Long{P}_{\mu\nu\rho\sigma}$ are given
in~\ref{2.35}, \ref{2.33}, the operator $\Tr{H}_{\mu\nu\rho\sigma}$ 
is defined in~\ref{224},
$$\Long{H}_{\mu\nu\rho\sigma}
\left(\frac{\partial}{\partial x}\right)
= \partial_\mu H_{\nu\rho\sigma}
\left(\frac{\partial}{\partial x}\right)
+\partial_\nu H_{\mu\rho\sigma}
\left(\frac{\partial}{\partial x}\right)
-\frac{2}{D}\delta_{\mu\nu}\partial_\lambda H_{\lambda\rho\sigma}
\left(\frac{\partial}{\partial x}\right),$$
where 
$$ H_{\mu\rho\sigma}
\left(\frac{\partial}{\partial x}\right) = 
\frac{2D^2-3D+2}{2D(D-1)}\partial_\mu\partial_\rho\partial_\sigma -
\frac{D-1}{2D}\left(\delta_{\mu\rho}\partial_\sigma+
\delta_{\mu\sigma}\partial_\rho\right)\Box - 
\frac{1}{2D(D-1)}\delta_{\rho\sigma}\partial_\mu\Box.$$
The coefficient in the second term is chosen in accordance with the Ward
identity~\ref{2.26}.

\section{Equivalence Conditions and Higher Green Functions.
The Propagators $\langle jj\rangle$ and $\langle TT\rangle$ on
Internal Lines}
\setcounter{equation}{0}

Let us consider higher Green functions which include the field $A_\mu(x)$
or $j_\mu(x)$ together with some different fields:
\begin{equation}\label{5.1}
G_\mu^A(x,\ldots) = \langle A_\mu(x)\Phi_1(x_1)\ldots\rangle,\quad
G_\mu^j(x,\ldots) = \langle j_\mu(x)\Phi_1(x_1)\ldots\rangle,
\end{equation}
where the dots stand for a certain set of fields $\Phi_i, i=1,2,\ldots$,
of arbitrary tensor structure. One can show~[1] that the two types of such
conformally invariant Green functions exist. The first type includes the
Green functions associated to irreducible representations $\tilde{Q}_A$
and $\Tr{Q}_j$, 
\begin{equation}\label{5.2}
G_{1\mu}^A(x,\ldots)=\langle A_\mu(x)\ldots\rangle_1,\quad
G_{1\mu}^j(x,\ldots)=\langle j_\mu(x)\ldots\rangle_1,
\end{equation}
while to the second type belong the Green functions associated to 
representations $\Long{Q}_A$ and $\tilde{Q}_j$:
\begin{equation}\label{5.3}
G_{2\mu}^A(x,\ldots)=\langle A_\mu(x)\ldots\rangle_2,\quad
G_{2\mu}^j(x,\ldots)=\langle j_\mu(x)\ldots\rangle_2.
\end{equation}
The Green functions $G_1^j$ and $G_2^A$ are associated with representations
which act on invariant subspaces $\Tr{M}_j$ and $\Long{M}_A$, see
section~3. So, the functions $G_1^j$ are transversal, while $G_2^A$ are
longitudinal:
\begin{equation}\label{5.4}
\partial_\mu^x G_{1\mu}^j(x,\ldots)=0,\quad
G_{2\mu}^A(x,\ldots)=\partial_\mu^x G(x,\ldots).
\end{equation}
Each of the Green functions~\ref{5.2},\ref{5.3} may be represented in terms
of partial wave expansion~[1], i.e., expressed through an infinite set of 
three-point functions of tensor fields 
$\Phi_s^l(x)=\Phi^l_{\mu_1\ldots\mu_s}(x)$ of dimension $l$, where $l,s$ 
run through all possible values. To prove~\ref{5.4} and other related 
statements, it is sufficient to demonstrate their validity for three-point 
functions.

Consider the functions~\ref{5.1} containing scalar fields. Introduce 
the invariant three-point functions
\begin{equation}\label{5.5}
C_\mu^{l,s}(x_1x_2x_3)=\langle\Phi^l_{\mu_1\ldots\mu_s}(x_1)
\varphi(x_2)j_\mu(x_3)\rangle,\quad
B_\mu^{l,s}(x_1x_2x_3)=\langle\Phi^l_{\mu_1\ldots\mu_s}(x_1)
\varphi(x_2)A_\mu(x_3)\rangle,
\end{equation}
where $\varphi(x)$ is a scalar field of dimension $d$. Both $C$- and 
$B$-functions fall into two types. Denote the functions of the 1st
type as $C_{1\mu}^{l,s},B_{1\mu}^{l,s}$, and of the 2nd --- as
$C_{2\mu}^{l,s},B_{2\mu}^{l,s}$. The explicit expressions for these functions
are found in~[1,2]. In particular, the functions $C_{1\mu}^{l,s}$ and 
$B_{2\mu}^{l,s}$, which satisfy~\ref{5.4}, have the form:
\begin{eqnarray} \label{5.6}
&& C_{1\mu}^{l,s}(x_1x_2x_3)\sim \biggl\{
\lambda_\mu^{x_3}(x_2x_1)\lambda_{\mu_1\ldots\mu_s}^{x_1}(x_2x_3) \nonumber\\
&&{}+\frac{(l-d)}{s(D-2-l+d+s)}
\frac{1}{x_{13}^2}\left[\sum_{k=1}^s g_{\mu\mu_k}(x_{13})
\lambda^{x_1}_{\mu_1\ldots\hat{\mu}_k\ldots\mu_s}(x_2x_3)-
\,\hbox{traces}\,\right]\biggr\}\nonumber\\
&&{}\times
\left(x_{12}^2\right)^{-\frac{l+d-s-D+2}{2}}
\left(x_{13}^2\right)^{-\frac{l-d-s+D-2}{2}}
\left(x_{23}^2\right)^{-\frac{d+s-l+D-2}{2}},
\end{eqnarray}
where 
$$\lambda_\mu^{x_3}(x_1x_2) = 
\frac{(x_{13})_\mu}{x_{13}^2}-
\frac{(x_{23})_\mu}{x_{23}^2}, \quad
\lambda^{x_1}_{\mu_1\ldots\mu_s}(x_2x_3) = 
\lambda^{x_1}_{\mu_1}(x_2x_3)\ldots
\lambda^{x_1}_{\mu_s}(x_2x_3) - \,\hbox{traces,}$$
$\hat{\mu}_k$ means that the index $\mu_k$ is omitted,
\begin{eqnarray} \label{5.7}
&& B_{2\mu}^{l,s}(x_1x_2x_3)\sim \biggl\{
\lambda_\mu^{x_3}(x_2x_1)\lambda_{\mu_1\ldots\mu_s}^{x_1}(x_2x_3) \nonumber\\
&&{}+\frac{1}{(l-d-s)}
\frac{1}{x_{13}^2}\left[\sum_{k=1}^s g_{\mu\mu_k}(x_{13})
\lambda^{x_1}_{\mu_1\ldots\hat{\mu}_k\ldots\mu_s}(x_2x_3)-
\,\hbox{traces}\,\right]\biggr\}\nonumber\\
&&{}\times
\left(x_{12}^2\right)^{-\frac{l+d-s}{2}}
\left(x_{13}^2\right)^{-\frac{l-d-s}{2}}
\left(x_{23}^2\right)^{\frac{l-d-s}{2}}.
\end{eqnarray}
One can check that these functions satisfy the conditions~\ref{5.4}
\begin{eqnarray} 
&&\label{5.8}
\partial_\mu^{x_3} C_{1\mu}^{l,s}(x_1x_2x_3)=0,\quad
\hbox{for all $l,s\ge1,$} \\
&&\label{5.9}
B_{2\mu}^{l,s}(x_1x_2x_3)\sim\partial_\mu^{x_3}\left[
\lambda_{\mu_1\ldots\mu_s}^{x_1}(x_2x_3)(x_{12}^2)^{-\frac{l+d-s}{2}}
\left(\frac{x_{23}^2}{x_{13}^2}\right)^{\frac{l-d-s}{2}}
\right].
\end{eqnarray}
Finally, the functions $C_{2\mu}^{l,s}$ and $B_{1\mu}^{l,s}$
may be derived from the expressions~\ref{5.6}, \ref{5.7}, if arbitrary
coefficients are inserted in front of the second terms in braces to
each of these expressions. Note that for $s=0$ no transversal function
$C_{1\mu}$ exists; each of the functions $\left.C_\mu^l\right|_{s=0}$
and $\left.B_\mu^l\right|_{s=0}$ has only one term.

For the Green functions $\langle A_\mu\ldots\rangle$ and 
$\langle j_\mu\ldots\rangle$, which are invariant under the new 
transformations~\ref{4.7} and~\ref{4.11}, the general equations may now
be easily written down. On account of~\ref{5.8},\ref{5.9} (for $s\ge1$)
one has for the three-point functions:
\begin{eqnarray} 
&& \label{5.10} 
B_\mu^{l,s}(x_1x_2x_3) = \langle\Phi_{\mu_1\ldots\mu_s}^{l,s}(x_1)
\varphi(x_2)A_\mu(x_3)\rangle \nonumber \\
&&\quad\quad\quad\quad{}=
\left(\delta_{\mu\nu}-
\frac{\partial_\mu^{x_3}\partial_\nu^{x_3}}{\Box_{x_3}}\right) 
B^{l,s}_{1\nu}(x_1x_2x_3)+B^{l,s}_{2\mu}(x_1x_2x_3) \\
&& \label{5.11} 
C_\mu^{l,s}(x_1x_2x_3) = \langle\Phi_{\mu_1\ldots\mu_s}^{l,s}(x_1)
\varphi(x_2)j_\mu(x_3)\rangle\nonumber \\
&&\quad\quad\quad\quad{} =
C^{l,s}_{1\mu}(x_1x_2x_3)+
\frac{\partial_\mu^{x_3}\partial_\nu^{x_3}}{\Box_{x_3}}
C^{l,s}_{2\mu}(x_1x_2x_3).
\end{eqnarray}
For the higher Green functions we obviously get
\begin{eqnarray} 
&& \label{5.12}
G^A_\mu(x,\ldots) = \langle A_\mu(x)\ldots\rangle = 
\left(\delta_{\mu\nu}-\frac{\partial_\mu\partial_\nu}{\Box}\right)
G_{1\nu}^A(x,\ldots) + \partial_\mu^x G(x,\ldots), \\
&& \label{5.13}
G^j_\mu(x,\ldots) = \langle j_\mu(x)\ldots\rangle = 
G_{1\mu}^j(x,\ldots)+
\frac{\partial_\mu\partial_\nu}{\Box} G_{2\nu}^j(x,\ldots).
\end{eqnarray}

A remarkable feature of conformal theory is the existence of relations
between Green functions of the current and those of the potential.
These relations are caused by the structure of representations of the 
conformal group, and are equivalent to Maxwell equations. Let us discuss it
in some detail.

The irreducible representations studied in section~3 are mutually related
by equivalence conditions~[12,13]
\begin{equation}\label{5.14}
\tilde{Q}_A\sim \Tr{Q}_j,\quad
\Long{Q}_A \sim \tilde{Q}_j.
\end{equation}
The invariant kernels~\ref{3.21} and~\ref{3.33} are the kernels of
intertwining operators which relate the fields $A_\mu$ and $j_\mu$. 
The first condition~\ref{5.14} links the Green functions~\ref{5.2},
and the second --- the functions~\ref{5.3}. Using the kernels~\ref{3.21},
we have:
\begin{eqnarray} 
\label{5.15} G_{1\mu}^j(x,\ldots) & = &
\int dy\,\Tr{\Delta}_{\mu\nu}(x-y)G_{1\nu}^A(y,\ldots), \\
\label{5.16} G_{2\mu}^j(x,\ldots) & = &\left.
\int dy\, \Delta^\epsilon_{\mu\nu}(x-y)G_{2\nu}^{A,\epsilon}(y,\ldots) 
\right|_{\epsilon=0},
\end{eqnarray}
where $\Delta^\epsilon_{\mu\nu}$ is the kernel~\ref{2.17}, singular at
$\epsilon=0$, and $G_{2\nu}^{A,\epsilon}$ is a regularized (see below)
longitudinal function~\ref{5.4}. These relations may be also rewritten
with the help of the kernels~\ref{3.33}:
\begin{eqnarray} 
\label{5.17} G_{1\mu}^A(x,\ldots) & = &\left.
\int dy\, D^\epsilon_{\mu\nu}(x-y)G_{1\nu}^{j,\epsilon}(y,\ldots)
\right|_{\epsilon=0}, \\
\label{5.18} G_{2\mu}^A(x,\ldots) & = &
\int dy\, \Long{D}_{\mu\nu}(x-y)G_{2\nu}^{A,\epsilon}(y,\ldots) ,
\end{eqnarray}
where $D_{\mu\nu}^\epsilon$ is the kernel~\ref{3.31}, singular at
$\epsilon=0$, and $G_{1\nu}^{j,\epsilon}$ is a regularized (see below)
transversal function, which satisfies~\ref{5.4}. Substituting
the explicit expression for the kernel $\Tr{\Delta}_{\mu\nu}$ into
Eq.\ref{5.15}, we get for $D=4$ the Maxwell equations for the Green
functions. Hence, the Maxwell equations express the equivalence conditions
for representations of the conformal group~[15,16].

As already mentioned, the higher Green functions can be represented in
the form of conformal partial wave expansions. 
To prove the relations~\ref{5.15}--\ref{5.18}, it is sufficient to 
demonstrate their validity for the invariant three-point functions
\begin{eqnarray} 
\label{5.19} C_{1\mu}^{l,s}(x_1x_2x_3) &= &\int dx_4 \,
\Tr{\Delta}_{\mu\nu}(x_{34}) B_{1\nu}^{l,s}(x_1x_2x_4), \\
\label{5.20} C_{2\mu}^{l,s}(x_1x_2x_3) &= &\left.\int dx_4 \,
\Delta^\epsilon_{\mu\nu}(x_{34}) B_{2\nu}^{l,s,\epsilon}(x_1x_2x_4)
\right|_{\epsilon=0}, \\
\label{5.21} B_{1\mu}^{l,s}(x_1x_2x_3) &= &\left.\int dx_4 \,
D^\epsilon_{\mu\nu}(x_{34}) C_{1\nu}^{l,s,\epsilon}(x_1x_2x_4)
\right|_{\epsilon=0} \\
\label{5.22} B_{2\mu}^{l,s}(x_1x_2x_3) &= &\int dx_4 \,
\Long{D}_{\mu\nu}(x_{34}) C_{2\nu}^{l,s}(x_1x_2x_4).
\end{eqnarray}
The calculation of these integrals is reviewed in~[1,2,4]. The conformal
regularization is used in~\ref{5.20} and~\ref{5.21}. The regularized
expressions may be derived by the substitution of anomalous dimensions
$l_j^\epsilon=D-1+\epsilon$, $l_A^\epsilon=1-\epsilon$ in place of
the canonical ones $l_j=D-1$, $l_A=1$. For the latter, it is sufficient
to introduce the factors
$(x_{12}^2)^{\epsilon/2}(x_{13}^2x_{23}^2)^{-\epsilon/2}$ and
$(x_{12}^2)^{-\epsilon/2}(x_{13}^2x_{23}^2)^{\epsilon/2}$ into
the expressions~\ref{5.6} and~\ref{5.7} correspondingly. Acting analogously,
one can derive the regularized expressions for the higher Green functions:
one just needs to represent them as conformal partial wave expansions in 
regularized three-point functions.

Let us remark that the equivalence conditions~\ref{5.20} and~\ref{5.21} may 
be rewritten without making use of regularized functions and kernels.
Indeed, it is sufficient to apply the realization obtained in the previous
section. We get instead of~\ref{5.20} and~\ref{5.21}
\begin{eqnarray} 
&& \label{5.23} 
\frac{\partial_\mu^{x_3}\partial_\nu^{x_3}}{\Box_{x_3}}
C_{2\nu}^{l,s}(x_1x_2x_3) = \int dx_4\, \Long{\Delta}_{\mu\nu}(x_{34})
B_{2\nu}^{l,s}(x_1x_2x_4), \\
&& \label{5.24} 
\left(\delta_{\mu\nu} -
\frac{\partial_\mu^{x_3}\partial_\nu^{x_3}}{\Box_{x_3}}\right)
B_{1\nu}^{l,s}(x_1x_2x_3) = \int dx_4\, \Tr{D}_{\mu\nu}(x_{34})
C_{1\nu}^{l,s}(x_1x_2x_4). 
\end{eqnarray}

Consider the conformally invariant integrals over internal photon
(or current) line
\begin{eqnarray} 
&& \label{5.25} \int dx\,dy\,
G_\mu^A(x,\ldots)\Delta_{\mu\nu}(x-y)G_\nu^A(y,\ldots) \\
&& \label{5.26} = \int dx\,dy\,
G_\mu^j(x,\ldots)D_{\mu\nu}(x-y)G_\nu^j(y,\ldots), 
\end{eqnarray}
where $G_\mu^A$ and $G_\mu^j$ are the Green functions~\ref{5.12}, \ref{5.13};
$\Delta_{\mu\nu}$ and $D_{\mu\nu}$ are the kernels~\ref{4.15} 
and~\ref{4.14}. Such integrals are encountered in conformally invariant
skeleton theory, as well as in exactly solvable $D$-dimensional models
considered in~[1--5]. The conformal invariance of the integrals~\ref{5.25}, 
\ref{5.26} is clear from the analysis of section~3. They represent
invariant contractions $(A,A)$ and $(j,j)$, see~\ref{3.36} and~\ref{3.24}.
Taking into account~\ref{3.17},\ref{3.19} and also \ref{3.28},\ref{3.32},
the integrals~\ref{5.25},\ref{5.26} can be rewritten through the regularized
expressions which are conformally invariant in a usual sense, i.e., with 
respect to transformations~\ref{3.6},\ref{3.11}:
\begin{eqnarray} \label{5.27}
&& \int G_\mu^A \Delta_{\mu\nu}^j G_\nu^A = 
\int dx\,dy\, G^A_{1\mu}(x,\ldots)
\Tr{\Delta}_{\mu\nu}(x-y) G^A_{1\nu}(y,\ldots) \nonumber \\
&&\qquad\quad{}+
\left. \int dx\,dy\, G^{A,\epsilon}_{2\mu}(x,\ldots)
\Delta^\epsilon_{\mu\nu}(x-y) G^{A,\epsilon}_{2\nu}(y,\ldots)
\right|_{\epsilon=0},
\end{eqnarray}
where $\Tr{\Delta}_{\mu\nu}$ and $\Delta^\epsilon_{\mu\nu}$
are the kernels~\ref{3.21}, see~\ref{3.7} and~\ref{2.17},
$G^A_{2\mu}$ is longitudinal,
\begin{eqnarray} \label{5.28}
&& \int G_\mu^j D_{\mu\nu}^A G_\nu^j = \left.
\int dx\,dy\, G^{j,\epsilon}_{1\mu}(x,\ldots)
D^{\epsilon}_{\mu\nu}(x-y) G^{j,\epsilon}_{1\nu}(y,\ldots) 
\right|_{\epsilon=0}\nonumber \\
&&\qquad\quad{}+
 \int dx\,dy\, G^j_{2\mu}(x,\ldots)
\Long{D}_{\mu\nu}(x-y) G^j_{2\nu}(y,\ldots),
\end{eqnarray}
where $D^\epsilon_{\mu\nu}$ and $\Long{D}_{\mu\nu}$
are the kernels~\ref{3.33}, see~\ref{3.31} and~\ref{3.8},
$G^j_{1\mu}$ is transversal. The left-hand sides of~\ref{5.27}
and~\ref{5.28} are equal (under the suitable renormalization of
the propagator $D_{\mu\nu}^A$) and may be expressed through invariant
kernels non-singular at $\epsilon=0$
\begin{eqnarray} \label{5.29}
&& \int G_\mu^A \Delta_{\mu\nu}^j G_\nu^A = 
 \int G_\mu^j D_{\mu\nu}^A G_\nu^j = \nonumber \\
&&\kern-20pt \int dx\,dy\, G^A_{1\mu}(x,\ldots)
\Tr{\Delta}_{\mu\nu}(x-y) G^A_{1\nu}(y,\ldots)+
 \int dx\,dy\, G^j_{2\mu}(x,\ldots)
\Long{D}_{\mu\nu}(x-y) G^j_{2\nu}(y,\ldots).\nonumber\\
\end{eqnarray}
All the conclusions concerning the relations~\ref{5.25}--\ref{5.29} are 
readily transferred to the case of invariant functions
$B_{1\mu}^{l,s}$, $B_{2\mu}^{l,s}$ and $C_{1\mu}^{l,s}$, $C_{2\mu}^{l,s}$
for $s\ge1$.

Consider the fields $h_{\mu\nu}$ and $T_{\mu\nu}$. Identically to the
case of the fields $A_\mu$, $j_\mu$, the irreducible 
representations~\ref{3.45} and~\ref{3.54} are pairwise equivalent~[12,13]:
\begin{equation}\label{5.30}
\tilde{Q}_h \sim \Tr{Q}_T,\quad
\Long{Q}_h \sim \tilde{Q}_T.
\end{equation} 
Accordingly, there are two types of invariant higher Green functions:
\begin{equation}\label{5.31}
G_{\mu\nu}^h(x,\ldots) = \langle h_{\mu\nu}(x)\ldots \rangle,\quad
G_{\mu\nu}^T(x,\ldots) = \langle T_{\mu\nu}(x)\ldots \rangle,
\end{equation}
where the dots stand for any sets of fields. The Green functions of
the first type transform by irreducible representations $\tilde{Q}_h$
and $\Tr{Q}_T$
\begin{equation}\label{5.32}
G_{1\mu\nu}^h(x,\ldots) = \langle h_{\mu\nu}(x)\ldots \rangle_1,\quad
G_{1\mu\nu}^T(x,\ldots) = \langle T_{\mu\nu}(x)\ldots \rangle_1,
\end{equation}
while those of the second type --- by representations $\Long{Q}_h$,
$\tilde{Q}_T$
\begin{equation}\label{5.33}
G_{2\mu\nu}^h(x,\ldots) = \langle h_{\mu\nu}(x)\ldots \rangle_2,\quad
G_{2\mu\nu}^T(x,\ldots) = \langle T_{\mu\nu}(x)\ldots \rangle_2.
\end{equation}
The function $G_{1\mu\nu}^T$ is transversal,
\begin{equation}\label{5.34}
\partial_\mu G_{1\mu\nu}^T(x,\ldots)=0,
\end{equation}
while the function $G_{2\mu\nu}^h$ is longitudinal
\begin{equation}\label{5.35}
G_{2\mu\nu}^h(x,\ldots) = \partial_\mu G_\nu^h(x,\ldots) +
\partial_\mu G_\mu^h(x,\ldots)-\frac{2}{D}\delta_{\mu\nu}
\partial_\lambda G_\lambda^h(x,\ldots),
\end{equation}
where $G_\mu^h(x,\ldots)$ is an invariant function of the vector field
$h_\mu(x)$, see~\ref{3.43}. 
For higher Green functions, which are invariant under the new 
transformations~\ref{4.7} and~\ref{4.11}, the general equations may now
be written using~\ref{5.35},\ref{5.36}:
\begin{eqnarray} 
\label{5.36} G_{\mu\nu}^h(x,\ldots)  & = &
\Tr{P}_{\mu\nu\rho\sigma}\left(\frac{\partial}{\partial x}\right)
G_{1\rho\sigma}^h(x,\ldots) +
G_{2\mu\nu}^h(x,\ldots), \\
\label{5.37} G_{\mu\nu}^T(x,\ldots)  & = &
G_{1\mu\nu}^T(x,\ldots) +
\Long{P}_{\mu\nu\rho\sigma}\left(\frac{\partial}{\partial x}\right)
G_{2\rho\sigma}^T(x,\ldots), 
\end{eqnarray}
where $\Tr{P}$ and $\Long{P}$ are the projection operators~\ref{2.35}
and~\ref{2.33}; $G_{1,2\,\mu\nu}^h$ and $G_{1,2\,\mu\nu}^T$
are invariant under the old transformations~\ref{3.39},\ref{3.41}.

The Green functions~\ref{5.31} may be expanded into infinite sets
of invariant three-point functions
\begin{equation}\label{5.38}
B^{l,s}_{\mu\nu}(x_1x_2x_3) = 
\langle \Phi_{\mu_1\ldots\mu_s}^l(x_1)\varphi(x_2)
h_{\mu\nu}(x_3)\rangle,\
C^{l,s}_{\mu\nu}(x_1x_2x_3) = 
\langle \Phi_{\mu_1\ldots\mu_s}^l(x_1)\varphi(x_2)
T_{\mu\nu}(x_3)\rangle.
\end{equation}
The explicit expressions for these functions may be found in~[1,2,5].
Exactly as in the case of the current, we have two types of 
functions~\ref{5.38} (for $s\ge2$):
\begin{equation}\label{5.39}
B^{l,s}_{1\mu\nu}(x_1x_2x_3),
C^{l,s}_{1\mu\nu}(x_1x_2x_3) \quad\hbox{and}\quad
B^{l,s}_{2\mu\nu}(x_1x_2x_3),
C^{l,s}_{2\mu\nu}(x_1x_2x_3),
\end{equation}
which correspond to the representations
$$\tilde{Q}_h,\Tr{Q}_T\quad\hbox{and}\quad\Long{Q}_h,\tilde{Q}_T.$$
Only one type survives when $s=0,1$~[1,2,5], namely,
$B^{l,s}_{2\mu\nu}$ and $C^{l,s}_{2\mu\nu}$. The function
$C^{l,s}_{1\mu\nu}$ is transversal, while $B^{l,s}_{2\mu\nu}$  is
longitudinal, see~\ref{5.34},\ref{5.35}.

Consider the equivalence conditions~\ref{5.30}. They mutually relate
each pair of functions~\ref{5.32} and~\ref{5.33}, as well as
three-point functions~\ref{5.39}. These relations may be written in
terms of kernels~\ref{3.46} or~\ref{3.42},\ref{3.50}:
\begin{eqnarray} 
\label{5.40} G_{1\mu\nu}^T(x,\ldots) & = &
\int dy\, \Tr{\Delta}_{\mu\nu\rho\sigma}(x-y)
G_{1\rho\sigma}^h(y,\ldots) \\
\label{5.41} G_{2\mu\nu}^T(x,\ldots) & = &
\left.\int dy\, \Delta^\epsilon_{\mu\nu\rho\sigma}(x-y)
G_{2\rho\sigma}^{h,\epsilon}(y,\ldots) \right|_{\epsilon=0},
\end{eqnarray}
where $\Tr{\Delta}_{\mu\nu\rho\sigma}$ is the invariant transversal
kernel~\ref{3.40}, and $\Delta^\epsilon_{\mu\nu\rho\sigma}$ is 
the kernel~\ref{2.30} singular in the $\epsilon=0$ limit. 
Let us remind that $G_{2\rho\sigma}^{h,\epsilon}$ is longitudinal
for $\epsilon=0$, see~\ref{5.35}, so that the integral~\ref{5.41} is
finite. The invariant regularization of the function $G_{2\rho\sigma}^h$
is provided via substitution of anomalous dimension $l_h^\epsilon=-\epsilon$
in place of canonical one $l_h=0$, as described above for the case of
potential $A_\mu$.

It is shown in~[1], see also~[17], that the relation~\ref{5.40} for
$D=4$ is equivalent to the equations of the linear conformal gravity.

The equalities~\ref{5.40},\ref{5.41} may be inverted and brought to
the form
\begin{eqnarray} 
\label{5.42} G_{1\mu\nu}^h(x,\ldots) & = &\left.
\int dy\, D^\epsilon_{\mu\nu\rho\sigma}(x-y)
G_{1\rho\sigma}^{T,\epsilon}(y,\ldots)\right|_{\epsilon=0} \\
\label{5.43} G_{2\mu\nu}^h(x,\ldots) & = &
\int dy\, \Long{D}_{\mu\nu\rho\sigma}(x-y)
G_{2\rho\sigma}^T(y,\ldots),
\end{eqnarray}
where $D^\epsilon_{\mu\nu\rho\sigma}$ is the ($\epsilon=0$)-singular 
kernel~\ref{3.50}, and $\Long{D}_{\mu\nu\rho\sigma}$ is the longitudinal
kernel~\ref{3.42}. The integral in~\ref{5.42} is finite since the function
$G_{1\rho\sigma}^{T,\epsilon}$ is transversal for $\epsilon=0$, and
the leading term $\sim1/\epsilon$ in the expansion of 
$D^\epsilon_{\mu\nu\rho\sigma}$ is longitudinal. Note that the 
relations~\ref{5.41} and~\ref{5.42} may be rewritten using non-singular
kernels if the realization of representations $\tilde{Q}_h$ and
$\tilde{Q}_T$ from section~4 is utilized.
\begin{eqnarray} 
\label{5.44} 
 \Long{P}_{\mu\nu\rho\sigma}\left(\frac{\partial}{\partial x}\right)
G_{2\rho\sigma}^T(x,\ldots) & = &
\int dy\, \Long{\Delta}_{\mu\nu\rho\sigma}(x-y)
G_{2\rho\sigma}^h(y,\ldots) \\
\label{5.45} 
 \Tr{P}_{\mu\nu\rho\sigma}\left(\frac{\partial}{\partial x}\right)
G_{1\rho\sigma}^h(x,\ldots) & = &
\int dy\, \Tr{D}_{\mu\nu\rho\sigma}(x-y)
G_{2\rho\sigma}^T(y,\ldots),
\end{eqnarray}
where $\Long{\Delta}_{\mu\nu\rho\sigma}$ and
$\Tr{D}_{\mu\nu\rho\sigma}$ are the finite kernels~\ref{2.28} and~\ref{3.52}.

The invariant three-point functions~\ref{5.39} are also related by 
equivalence conditions. The relations of the type~\ref{5.40}--\ref{5.45}
for these functions may be proved by direct calculations. The technical
details of the such calculations may be found in~[1], see also~[2,5].

It remains to discuss invariant integrals over internal lines 
corresponding to the fields $h_{\mu\nu}$ and $T_{\mu\nu}$. A literal
quotation of the arguments concerning the derivation of 
equations~\ref{5.25}--\ref{5.29} in the case of the fields $A_\mu$, 
$j_\mu$ is sufficient, and will not be reproduced here. The only relevant
comment is that the properties of invariant 
forms~\ref{3.38},\ref{3.48},\ref{3.57} and the 
expressions~\ref{4.19},\ref{4.20} for the propagators $\langle hh\rangle$
and $\langle TT\rangle$ are applied in this case. 

\section{Irreducible Components of the Current and the
Ene\-r\-gy\--Mo\-men\-tum Tensor}
\setcounter{equation}{0}
The two types of Green functions for the current were analyzed in the 
previous section. These functions correspond to the pair 
of irreducible representations $\Tr{Q}_j$ and $\tilde{Q}_j$. It proves
helpful to introduce two fields: $\Tr{j}_\mu(x)$ and $\tilde{j}_\mu(x)$.
The total Euclidean current $j_\mu(x)$ transforms by the irreducible
representation $\Tr{Q}_j \oplus \tilde{Q}_j$ and may be written as
\begin{equation}\label{6.1}
j_\mu(x) = \Tr{j}_\mu(x)+\tilde{j}_\mu(x).
\end{equation}
Accordingly, one has a pair of propagators:
\begin{equation}\label{6.2}
\langle\Tr{j}_\mu(x_1)\Tr{j}_\nu(x_2)\rangle\quad\hbox{and}\quad
\langle\tilde{j}_\mu(x_1)\tilde{j}_\nu(x_2)\rangle.
\end{equation}
All the Green functions $\langle\Tr{j}_\mu\ldots\rangle$ of the
current $\Tr{j}$ are transversal, and its propagator has the form
\begin{equation}\label{6.3}
\Tr{\Delta}_{\mu\nu}(x_{12}) = 
\langle\Tr{j}_\mu(x_1)\Tr{j}_\nu(x_2)\rangle \sim
\left(\delta_{\mu\nu}\Box-\partial_\mu\partial_\nu\right)
\Box^{\frac{D-4}{2}}\delta(x_{12}).
\end{equation}

The Green functions of the current $\tilde{j}_\mu$ depend on the choice
of realization of the representation $\tilde{Q} _j$. Choosing different
representatives in the equivalence class, see sections~3,4, one can
obtain different realizations of the Green functions $\langle\tilde{j}_\mu
\ldots\rangle$. In section~4 we have examined a special realization
of the representation $\tilde{Q}_j$, where the current $\tilde{j}_\mu$
is longitudinal. The propagator of the current in this realization
takes the form
\begin{equation}\label{6.4}
\langle\tilde{j}_\mu(x_1)\tilde{j}_\nu(x_2)\rangle = C_j
\partial_\mu\partial_\nu \Box^{\frac{D-4}{2}}\delta(x_{12}).
\end{equation}
Note that the relevant conformal transformations are non-local and
may differ from the transformations of the current $\Tr{j}_\mu$,
see~\ref{4.11}. The propagator of the total current equals to the
sum of terms~\ref{6.3} and~\ref{6.4}. The other realization of 
the representation $\tilde{Q}_j$ has also been studied; the transformations
in that one are local and coincide with the transformations of
the current $\Tr{j}_\mu$. In the latter realization the propagator of
the current demands a regularization (section~2):
\begin{eqnarray}\label{6.5}
&& \langle\tilde{j}_\mu(x_1)\tilde{j}_\nu(x_2)\rangle = 
\Delta^\epsilon_{\mu\nu}(x_{12}) = 
C_j\partial_\mu\partial_\nu\Box^{\frac{D-4}{2}}\delta(x_{12})
\nonumber \\
&&\qquad\qquad{}
+A(\epsilon)\left(\delta_{\mu\nu}\Box-\partial_\mu\partial_\nu\right)
\frac{1}{(x_{12}^2)^{D-2+\epsilon}}+O(\epsilon).
\end{eqnarray}
However this is the equivalence class $\left[j_\mu\right]$ and not
the current $\tilde{j}_\mu$ by itself that has the physical meaning.
The framework of conformal theory is constructed in such a manner
that the transformations inside an equivalence class do not effect
the values of the conformally invariant contractions~\ref{5.25},\ref{5.26}.
The latter ones may, in particular, be brought to the 
form~\ref{5.27},\ref{5.28} or~\ref{5.29}. This property of contractions
was discussed in section~3 in connection with the specification of
invariant forms in the spaces of representations $\tilde{Q}_j$,
$\Long{Q}_A$, see~\ref{3.19} and \ref{3.28}. Hence 
one is free use either realization:~\ref{6.4} or~\ref{6.5},
depending on what is actually needed. The transversal component 
in~\ref{6.5} is divergent at $\epsilon=0$ and does not contribute to 
conformally invariant graphs. However, the realization~\ref{6.5} is 
more convenient for technical purposes, as demonstrated in~[2--5], 
see~[1] for more details.
 
All the above is also valid for the field $A_\mu(x)$, which transforms
by the irreducible representation $\tilde{Q}_A \oplus \Long{Q}_A$.
Introducing irreducible fields $\tilde{A}_\mu(x)$ and $\Long{A}_\mu(x)$,
rewrite the field $A_\mu(x)$ as the sum
\begin{equation}\label{6.6}
A_\mu(x) = \tilde{A}_\mu(x)+\Long{A}_\mu(x).
\end{equation}
All the Green functions of the field $\Long{A}_\mu$ are longitudinal.
Its propagator has the form
\begin{equation}\label{67}
\langle\Long{A}_\mu(x_1)\Long{A}_\nu(x_2)\rangle=
\Long{D}_{\mu\nu}(x_{12})\sim\partial_\mu\partial_\nu\ln x_{12}^2.
\end{equation}
The Green functions of the field $\tilde{A}_\mu$ depend on the choice of 
realization of the representation $\tilde{Q}_A$. Different realizations 
correspond to different representatives of the equivalence class 
$\left[A_\mu\right]$. In section~4 we discussed a realization in which
the propagator 
$\langle\tilde{A}_\mu\tilde{A}_\nu\rangle$ is transversal:
\begin{equation}\label{6.7}
\langle\tilde{A}_\mu(x_1)\tilde{A}_\nu(x_2)\rangle \sim
\left(\delta_{\mu\nu}\Box-\partial_\mu\partial_\nu\right)\ln x_{12}^2.
\end{equation}
The conformal transformation in this realization are non-local and
differ from the transformations of longitudinal field $\Long{A}_\mu$,
see~\ref{4.4}--\ref{4.6}. Another realization, in which the conformal
transformations are local and coincide with the transformations of the
field $\Long{A}_\mu$, is also studied in section~3. In the latter case,
the propagator $\langle\tilde{A}_\mu\tilde{A}_\nu\rangle$ demands
a regularization:
\begin{equation}\label{6.8}
\langle\tilde{A}_\mu(x_1)\tilde{A}_\nu(x_2)\rangle = 
D_{\mu\nu}^\epsilon(x_{12}) \sim \frac{1}{\epsilon}
(x_{12}^2)^{-1+\epsilon}g_{\mu\nu}(x_{12}).
\end{equation}
Unlike~\ref{6.7}, it has a longitudinal component which is singular
at $\epsilon=0$ and does not contribute to conformally invariant graphs,
see~\ref{5.24},\ref{5.28} and~\ref{5.29}. Technically, the 
realization~\ref{6.8} is more useful.

Thus, henceforth to the end of this paper we use the realization~\ref{6.5} 
and~\ref{6.8}. The equivalence conditions for the representations~\ref{5.14}
may be written in the form of operator relations~[1]: in the transversal 
sector
\begin{equation}\label{6.9}
\Tr{j}_\mu(x) = \int dy\, \Tr{\Delta}_{\mu\nu}(x-y)\tilde{A}_\nu(y),\
\tilde{A}_\mu(x) = \left.\int dy\, D^\epsilon_{\mu\nu}(x-y) 
j^{{\rm tr}, \epsilon}_\nu(y)\right|_{\epsilon=0},
\end{equation}
and in the longitudinal sector
\begin{equation}\label{6.10}
\tilde{j}_\mu(x) = \left.\int dy\, \Delta^\epsilon_{\mu\nu}(x-y) 
A^{{\rm long},\epsilon}_\nu(y)\right|_{\epsilon=0},\ 
\Long{A}_\mu(x) = \int dy\, \Long{D}_{\mu\nu}(x-y) \tilde{j}_\nu(y).
\end{equation}
Here the relations between Euclidean quantum fields are considered
as a condensed form of analogous relations for all the Green functions,
see~\ref{5.15}--\ref{5.18}. The regularized fields 
$A^{{\rm long},\epsilon}_\mu$ and $j^{{\rm td},\epsilon}_\mu$
signify that the regularization has been introduced to the Green functions, 
see section~5. We remind that the equations~\ref{6.9} in $D=4$ are
equivalent to the Maxwell equations~[1].


The physical meaning of expansions of Euclidean fields
$j_\mu$ and $A_\mu$ into pairs of irreducible components
may be commented in the following manner.
Consider the Euclidean fields $j_\mu(x)$ and $T_{\mu\nu}(x)$.
The Green functions $\langle j_\mu\ldots\rangle$ and $\langle
T_{\mu\nu}\ldots\rangle$ are the Euclidean analogues of $T$-ordered   
vacuum expectation values in Minkowski space. Here we treat the
Euclidean fields $j_\mu(x)$ and $T_{\mu\nu}(x)$ as the symbolic
notation for the complete sets of Green functions 
$\langle j_\mu\ldots\rangle$ and $\langle T_{\mu\nu}\ldots\rangle$.
Correspondingly, the derivatives of the Euclidean fields
$\partial_\mu j_\mu(x)$ and $\partial_\mu T_{\mu\nu}(x)$ denote
the derivatives of Green functions
$\partial_\mu\langle j_\mu\ldots\rangle$ and $\partial_\mu\langle 
T_{\mu\nu}\ldots\rangle$. Calculating these derivatives, one
encounters the two types of terms of different nature. Consider
those terms on an example of the conserved current in Minkowski space.
One gets for the propagator of the current:
$$ \partial_\mu \vac{T\left\{j_\mu(x)j_\nu(0)\right\}} = 
\delta(x^0)\vac{\left[j_0(x),j_\nu(0)\right]}+
\vac{T\left\{\partial_\mu j_\mu(x)j_\nu(0)\right\}}.$$
The second term vanishes due to the conservation law
$$\partial_\mu j_\mu^{\rm Mink}(x)=0.$$
To ensure the covariance of the $T$-ordered average, one should add
quasilocal terms to the first term of the expression. The form of the total
contribution of these terms and the commutator is imposed by conformal
invariance, and reads
$$ \partial_\mu \vac{T\left\{j_\mu(x)j_\nu(0)\right\}} = 
C_j\partial_\nu\Box^{\frac{D-2}{2}}\delta(x).$$
In Euclidean conformal theory, one associates the above pair of contributions
with the irreducible components of the Euclidean current $j_\mu(x)$,
so that
$$\partial_\mu j_\mu(x) = \partial_\mu\tilde{j}_\mu(x),\quad
\partial_\mu\Tr{j}_\mu(x)=0.$$
In particular, the total propagator of the current
may be represented as:
\begin{eqnarray*}
&& \langle j_\mu(x_1)j_\nu(x_2)\rangle = 
 \langle \tilde{j}_\mu(x_1)\tilde{j}_\nu(x_2)\rangle +
 \langle \Tr{j}_\mu(x_1)\Tr{j}_\nu(x_2)\rangle, \\
&& \partial_\mu\langle j_\mu(x_1)j_\nu(x_2)\rangle = 
 \partial_\mu\langle \tilde{j}_\mu(x_1)\tilde{j}_\nu(x_2)\rangle  = 
 C_j\partial_\nu\Box^{\frac{D-2}{2}}\delta(x_{12}),\\
&& \partial_\mu\langle \Tr{j}_\mu(x_1)\Tr{j}_\nu(x_2)\rangle = 0.
\end{eqnarray*}
Thus, the two irreducible components
$\tilde{j}_\mu$ and $\Tr{j}_\mu$ have different physical meaning, and hence
the different group-theoretic structure.

Only the current
$\Tr{j}_\mu(x)$, but not $\tilde{j}_\mu(x)$, induces a non-trivial 
contribution to the electromagnetic interaction. The Green functions
of the current $\tilde{j}_\mu$ satisfy non-trivial Ward identities and
contain the information on the (postulated) commutation relations of
the total current:
$$ \left[ j_0(x),j_k(0)\right]_{x^0=0},
\left[ j_0(x),\varphi(0)\right]_{x^0=0},\ldots.$$
As shown in [1,2], all the Green functions $\langle\tilde{j}_\mu\ldots
\rangle$ are uniquely determined by the condition of conformal invariance
and by the Ward identities. Resultantly, the operator product
expansions $\tilde{j}_\mu(x)\varphi(0)$ have the form~[1,2]
$$\tilde\j_\mu(x)\varphi(0) = \sum_s \left[P_s\right],$$
where $P_s$ are the tensor fields of rank $s$ and dimension $d+s$.
This result does not depend on the type of interaction and is 
intrinsically due by conformal symmetry and the contributions of equal-time 
commutators. The dynamical models are defined~[1,2] by the definition
of commutators $\left[j_\mu(x),j_\nu(0)\right]_{x^0=0}$.
In Euclidean version of the theory, this commutator is
determined by the type of the operator product expansion
$\tilde{j}_\mu(x)\tilde{j}_\nu(0)$. 
The expansion $\tilde{j}_\mu(x)\tilde{j}_\nu(0) = 
\left[C_j\right]+\left[P_j\right]+\ldots$ was considered in~[1,2],
while in~[4] the models with $C_j\ne 0$,
$P_j(x)=0$ were examined.

One should remark that since the current $\tilde{j}_\mu(x)$ arises as
a representative of an equivalence class $\left\{\tilde{j}_\mu\right\}
\subset \tilde{M}_j = M_j/\Tr{M}_j$,
the transversal parts of the Green
functions $\langle\tilde{j}\ldots\rangle$ may be redefined by performing
a different choice of representatives. Particularly, in the non-local 
realization of conformal transformations, considered
above, these Green functions are longitudinal. This realization
is useful for conformal QED. However, it is essential that the local
realization of conformal transformations of the current $\tilde{j}_\mu$,
is needed for the analysis of the operator
product expansions $\tilde{j}_\nu\varphi$ and $\tilde{j}_\mu
\tilde{j}_\nu$.
In a local realization, the Green 
functions $\langle\tilde{j}_\mu\ldots\rangle$ have quite definite 
transversal parts, which do not contribute to the electromagnetic interaction
since an irreducible component $\tilde{j}_\mu$ of the total current
only appears in contractions of the type $\int dx\,
\tilde{j}_\mu(x)\Long{A}_\mu(x)$.
The interaction with the irreducible field 
$\tilde{A}_\mu$ is caused by the component $\Tr{j}_\mu$ of the total
current, and has the form 
$\int dx\,\Tr{j}_\mu(x)\tilde{A}_\mu(x)$.


Let us consider the structure of the Hilbert space of conformal theory more
comprehensively. As shown in~[1], see also~[2], the two types of conformal
currents
\begin{equation}\label{6.11}
\tilde{j}_\mu(x)\quad\hbox{and}\quad\Tr{j}_\mu(x)
\end{equation}
may be associated with the two mutually orthogonal sectors of the Hilbert 
space
\begin{equation}\label{6.12}
\tilde{H}\oplus H_0,
\end{equation}
where $\tilde{H}$ is generated by the states
\begin{equation}\label{613}
\tilde{j}_\mu(x_1)\varphi(x_2)\mid0\rangle,\
\tilde{j}_\mu(x_1)\tilde{j}_\nu(x_2)\varphi(x_3)\mid0\rangle,\ldots,
\end{equation}
and $H_0$ includes analogous states of the current $\Tr{j}_\mu(x)$.
The orthogonality of the spaces $\tilde{H}$ and $H_0$ means the vanishing
of the Green functions
\begin{equation}\label{6.13}
\langle\Tr{j}_\mu(x_1)\tilde{j}_\nu(x_2)\rangle=0,\quad
\langle\varphi(x_1)\Tr{j}_\mu(x_2)\tilde{j}_\nu(x_3)
\varphi^\dagger\rangle=0.
\end{equation}
Due to the equivalence condition~\ref{6.9} the subspace $H_0$ contains 
nothing but electromagnetic degrees of freedom. The arising of non-zero
current $\Tr{j}_\mu$ necessarily begets the electromagnetic interaction.
If we are engaged in the analysis of non-electromagnetic interaction,
i.e., are interested in the states of subspace $\tilde{H}$, the problem of
separation of the current $\tilde{j}_\mu$ from the total current~\ref{6.1}
emerges.

The latter leads to the following situation. The conformal symmetry arises 
as a non-perturbative effect. The conclusions of conformal theory cannot have 
any analogues in perturbation theory. Moreover, the conformal symmetry may 
occur in a special class of models, not necessarily lagrangean. 
Given the structure of the Hilbert space described above, the original
presence of gauge interaction in conformal models would be the most
natural conjecture. Under that, both irreducible components~\ref{6.11}
contribute to the total current.
A ``true'' conformal theory must
include the complete current~\ref{6.1}, and hence, due to~\ref{6.9},
the gauge field $\tilde{A}_\mu$ as well. If one is going to examine an 
approximate model without gauge interactions, the solution is to be 
looked for in the restricted class of Green functions $\langle j_\mu\ldots
\rangle$, those corresponding to the irreducible representation 
$\tilde{Q}_j$. This leads to certain restrictions on higher Green functions 
which guarantee the irreducibility:
\begin{equation}\label{6.14}
j_\mu(x)=\tilde{j}_\mu(x),\quad \Tr{j}_\mu(x) = 0 \quad 
\hbox{on $\tilde{H}$.}
\end{equation} 
Such restrictions were studied to a fair extent in~[1,2], see also~[4].

These works deal with the class of non-gauge models in $D$-dimensional space,
which are analogous to two-dimensional conformal models. The irreducibility
condition for the current~\ref{6.14} is written in the following form~[1,2,4]
\begin{equation}\label{6.15}
\int dy\,dz\, B_{1\mu}^{l,s}(xyz)\langle j_\mu(z)\varphi(y)\ldots\rangle=0
\quad\hbox{for all $l,s$,}
\end{equation}
where $B_{1\mu}^{l,s}$ are the invariant three-point functions introduced in 
section~5. A theory supplied with such a condition is non-trivial if the 
operator product expansion $j_\mu(x_1)j_\nu(x_2)$, where 
$j_\mu(x)=\tilde{j}_\mu(x)$ is the irreducible current, includes the anomalous 
terms $\left[C_j\right]$ and $\left[P_j\right]$, see Introduction. The 
conditions~\ref{6.15} allow one to calculate all the Green functions of the 
current from anomalous~[3] Ward identities.

All the above is easily generalized to the case of the energy-momentum tensor 
and the metric field. One introduces a pair of tensor fields
\begin{equation}\label{6.16}
\Tr{T}_{\mu\nu}(x)\quad\hbox{and}\quad\tilde{T}_{\mu\nu}(x),
\end{equation}
transforming by irreducible representations $\Tr{Q}_T$ and $\tilde{Q}_T$.
All the Green functions of the field $\Tr{T}_{\mu\nu}$ are transversal,
and the propagator is given by the expression~\ref{3.40}. The Green functions
of the field $\tilde{T}_{\mu\nu}$ solely depend on the choice of realization 
of the representation $\tilde{Q}_T$. In the non-local realization of the 
section~4, the propagator $\langle\tilde{T}\tilde{T}\rangle$ has the 
form~\ref{2.28}. The total propagator of the field 
\begin{equation}\label{6.17}
T_{\mu\nu}(x)=\Tr{T}_{\mu\nu}(x)+\tilde{T}_{\mu\nu}(x)
\end{equation}
in this realization reads
\begin{equation}\label{6.18}
\langle T_{\mu\nu}(x_1)T_{\rho\sigma}(x_2)\rangle = 
\Tr{\Delta}_{\mu\nu\rho\sigma}(x_{12})+
\Long{\Delta}_{\mu\nu\rho\sigma}(x_{12})
\end{equation}
and is given by Eq.\ref{4.20}. In the local realization of the representation 
$\tilde{Q}_T$ the propagator of the field $\tilde{T}_{\mu\nu}$ is given by 
the regularized expression~\ref{2.30}. Identically to the case of 
electromagnetic interaction, the conformally invariant graphs do not depend 
on the choice of realization of the representation $\tilde{Q}_T$. This is 
obvious from the inspection of invariant forms on the spaces of 
representations $\tilde{Q}_T,\Long{Q}_h$, see section~3. (The choice of 
representatives in the equivalence classes $\left[T_{\mu\nu}\right]$
and $\left[h_{\mu\nu}\right]$ does not alter the values of the forms.)
Analogously, the metric tensor
\begin{equation}\label{6.19}
h_{\mu\nu}(x) = \tilde{h}_{\mu\nu}(x)+\Long{h}_{\mu\nu}(x)
\end{equation}
also transforms by the reducible representation 
$\tilde{Q}_h\oplus\Long{Q}_h$. The propagator of the irreducible field
$\Long{h}_{\mu\nu}$
is longitudinal, see~\ref{3.42}, while the propagator of the field 
$\tilde{h}_{\mu\nu}$ has a transversal part. In the realization of section~4 
the total propagator of the field $h_{\mu\nu}$ is given by the expression,
see~`\ref{4.19}
\begin{equation}\label{6.20}
\langle h_{\mu\nu}(x_1)h_{\rho\sigma}(x_2)\rangle=
\Tr{D}_{\mu\nu\rho\sigma}(x_{12})+
\Long{D}_{\mu\nu\rho\sigma}(x_{12}).
\end{equation}
In the local realization of the representation $\tilde{Q}_h$ the propagator
of the field $\tilde{h}_{\mu\nu}$ is given by the regularized 
expression~\ref{3.50}.

Consider the equivalence conditions for the representations~\ref{5.30}.
In the operator notation they read:
\begin{eqnarray} 
&& \label{6.21} \Tr{T}_{\mu\nu}(x) = \int dy\,
\Tr{\Delta}_{\mu\nu\rho\sigma}(x-y)\tilde{h}_{\rho\sigma}(y),\
\tilde{h}_{\mu\nu}(x) = \left.
\int dy\, D^\epsilon_{\mu\nu\rho\sigma}(x-y)
T^{{\rm tr},\epsilon}_{\rho\sigma}(y)\right|_{\epsilon=0}, \\
&& \label{6.22} \tilde{T}_{\mu\nu}(x) =\left. \int dy\,
\Delta^\epsilon_{\mu\nu\rho\sigma}(x-y)
h^{{\rm long},\epsilon}_{\rho\sigma}(y)\right|_{\epsilon=0},\
\Long{h}_{\mu\nu}(x) = \int dy\, \Long{D}_{\mu\nu\rho\sigma}(x-y)
\tilde{T}_{\rho\sigma}(y). \nonumber \\
\end{eqnarray}
The kernels $\Tr{\Delta}_{\mu\nu\rho\sigma}$ and 
$D^\epsilon_{\mu\nu\rho\sigma}$ are given by the expressions~(3.40)
and~(3.50), while
the kernels $\Delta^\epsilon_{\mu\nu\rho\sigma}$ and 
$\Long{D}_{\mu\nu\rho\sigma}$ --- by the expressions~(2.30) and~(3.42).

The regularization of the fields $\Tr{T}_{\rho\sigma}$
and $\Long{h}_{\rho\sigma}$ means the regularization of the Green functions, 
see section~5. Let us remind that when $D=4$, the equations~\ref{6.21} are 
equivalent to the equations of linear conformal gravity.

The states of the Hilbert space generated by the fields~\ref{6.16} also form 
a pair of orthogonal subspaces $H_0$ and $\tilde{H}$. Due to the equivalence 
conditions~\ref{6.21} the subspace $H_0$ contains solely gravitational 
degrees of freedom. The field $\Tr{T}_{\mu\nu}$ necessarily begets a 
gravitational interaction. Recalling the arguments presented above, we 
conclude that the theory without gravitational interaction must include the 
irreducible tensor $\tilde{T}_{\mu\nu}$ only, and satisfy the condition
\begin{equation}\label{6.23}
T_{\mu\nu}(x)=\tilde{T}_{\mu\nu}(x),\quad
\Tr{T}_{\mu\nu}(x)=0 \quad \hbox{on $\tilde{H}$.}
\end{equation}
As shown in~[1,2], see also~[5], these conditions may be set up as the 
following conditions on higher Green functions
\begin{equation}\label{6.24}
\int dy\,dz\, B^{l,s}_{1\mu\nu}(xyz)
\langle T_{\mu\nu}(z)\varphi(y)\ldots\rangle=0 \quad
\hbox{for all $l,s$,}
\end{equation}
where $B^{l,s}_{1\mu\nu}$ are the invariant functions discussed in section~5.
A theory supplied with such a condition is non-trivial if the 
operator product expansion $T_{\mu\nu}(x_1)T_{\rho\sigma}(x_2)$, where 
$T_{\mu\nu}(x)=\tilde{T}_{\mu\nu}(x)$ is the irreducible field, includes 
the anomalous terms $\left[C_T\right]$ and $\left[P_T\right]$, see 
Introduction. 

Thus the Green functions of the irreducible fields
$$\langle\tilde{j}_\mu\varphi\ldots\rangle,\quad 
\langle\tilde{T}_{\mu\nu}\varphi\ldots\rangle $$
satisfy the conditions~\ref{6.15} and~\ref{6.24}, and are {\bf uniquely}
determined by anomalous Ward identities for any space dimension (either
even or odd~[1,2,5]). Such a theory acquires analogues of null-vectors, each
defining an exactly solvable model~[1,2,4,5]. Note that the current and
the energy-momentum tensor of two-dimensional theory are analogous to the 
fields $\tilde{j}_\mu$ and $\tilde{T}_{\mu\nu}$, while the fields 
$\Tr{j}_\mu$ and $\Tr{T}_{\mu\nu}$ have no analogues when $D=2$. The 
propagator of the energy-momentum tensor in $D=2$ has the form~\ref{1.7}.
Using the identity~\ref{1.6} one can bring the propagator to the form which 
follows from~\ref{2.28},\ref{2.29} in $D=2$:
$$\left.\langle T_{\mu\nu}(x_1)T_{\rho\sigma}(x_2)\right|_{D=2}=
\partial_\mu\Delta_{\nu\rho\sigma}(x_{12})+\partial_\nu
\Delta_{\mu\rho\sigma}(x_{12})-\delta_{\mu\nu}\partial_\lambda
\Delta_{\lambda\rho\sigma}(x_{12}),$$
where
$$\left.\Delta_{\mu\rho\sigma}\right|_{D=2}
\sim\left[4\partial_\mu\partial_\rho\partial_\sigma-
\left(\delta_{\mu\rho}\partial_\sigma+\delta_{\mu\sigma}\partial_\rho\right)
\Box-\delta_{\rho\sigma}\partial_\mu\Box\right]\ln x^2.$$
Therefore an irreducible representation (of the 6-parametric conformal group 
in $D=2$) corresponding to it, is analogous to the representation $\tilde{Q}$ 
in $D$-dimensional theory. The theory defined by the conditions~\ref{6.23} 
and~\ref{6.24} is a natural candidate to the generalization of 
two-dimensional conformal theory.

\newpage 
\section*{Acknowledgments}
\label{Acknowledgments}
E.S.F. thanks Phillip A. Griffiths and the Institute for Advanced
Study for warm hospitality and Stephen Adler for 
interesting discussions. This work is partially supported 
by RFBR grants No. 96--02--18966
and No. 96--15--96463.

\newpage

\def\hf{\hfil\break}

\section*{Bibliography}

\begin{enumerate}
\item E.S.Fradkin and M.Ya.Palchik, \hf Conformal Quantum Field Theory
      in $D$-Dimensions, Kluwer Acad. Publ., 1996
\item E.S.Fradkin and M.Ya.Palchik, Ann. of Phys. 249(1996)44.
\item E.S.Fradkin, M.Ya.Palchik, and V.N.Zaikin, Phys.Rev. D53(1996)7345.
\item E.S.Fradkin and M.Ya.Palchik,
      preprint, 1C/96/21, Triest, 1996.
\item E.S.Fradkin and M.Ya.Palchik,
      preprint, 1C/96/22, Triest, 1996.
\item A.A.Belavin, A.M.Polyakov and A.B.Zamolodchikov, \hf Nucl.Phys.
      B241(1984)333.
\item D.Friden, Z.Qiu and S.Shenker,
      Phys.Rev.Lett. 52(1984)1575,\hf  Phys.Lett. 151B(1985)37.
\item V.G.Knizhnik and A.B.Zamolodchikov,\hf Nucl.Phys. B247(1984)33.
\item H.~Osborn and A.~C.~Petkou, Ann. Phys. (N.Y.) {\bf 231} (1994) 311.
\item J.~Erdmenger and H.~Osborn,
      Nucl.Phys., B483 (1997) 431.
\item I.M.Gel'fand and G.E.Shilov, Generalized Functions,
      v.1 (Academic Press, New York, 1964).
\item V.K.Dobrev, G.Mack, V.B.Petkova, S.G.Petrova and I.T.Todorov,
      Lecture Notes in Physics, v.63 (Springer-Verlag, 1977).
\item A.U.Klimyk and A.M.Gavrilik, Matrics Elements and Klebsh-Gordan
      Coefficients of Group Representations (Naukova Dumka, Kiev, 1979).
\item M.Ya.Palchik, J.Phys. 16(1983)1523.
\item A.A.Kozhevnikov, M.Ya.Palchik and A.A.Pomeranskii,\hf  Yadernaya
      Fizika, 37(1983)481.
\item E.S.Fradkin, A.A.Kozhevnikov, M.Ya.Palchik and A.A.Pomeranskii,
      Comm. Math. Phys. 91(1983)529.
\item E.S.Fradkin and M.Ya.Palchik, Class.Quantum Grav. 1(1984)131.
\end{enumerate}

\end{document}